\documentclass[twocolumn]{aastex63}

\received{...}
\revised{...}
\accepted{...}
\submitjournal{ApJ}

\usepackage{rotating}

\shorttitle{Long-term mid-IR monitoring of SNe}
\shortauthors{Szalai et al.}

\begin{document}

\title{Spitzer's Last Look at Extragalactic Explosions: Long-Term Evolution of Interacting Supernovae}

\correspondingauthor{Tam\'as Szalai}
\email{szaszi@titan.physx.u-szeged.hu}

\author[0000-0003-4610-1117]{Tam\'as Szalai}
\affiliation{Department of Optics and Quantum Electronics, Institute of Physics, University of Szeged, D\'om t\'er 9, Szeged, 6720 Hungary}
\affiliation{Konkoly Observatory, Research Centre for Astronomy and Earth Sciences, H-1121 Budapest, Konkoly Thege Miklós út 15-17,Hungary}

\author[0000-0003-2238-1572]{Ori D. Fox}
\affiliation{Space Telescope Science Institute, 3700 San Martin Drive, Baltimore, MD 21218, USA}

\author[0000-0001-8403-8548]{Richard G. Arendt}
\affiliation{Code 665, NASA/GSFC, 8800 Greenbelt Road, Greenbelt, MD 20771, USA}
\affiliation{CRESST II/UMBC, USA; Richard.G.Arendt@nasa.gov}

\author[0000-0001-8033-1181]{Eli Dwek}
\affiliation{Code 665, NASA/GSFC, 8800 Greenbelt Road, Greenbelt, MD 20771, USA}

\author[0000-0003-0123-0062]{Jennifer E. Andrews}
\affiliation{Steward Observatory, University of Arizona, 933 North Cherry Avenue, Tucson, AZ 85721-0065, USA}

\author[0000-0002-0141-7436]{Geoffrey C. Clayton}
\affiliation{Department of Physics \& Astronomy, Louisiana State University, Baton Rouge, LA 70803, USA}

\author[0000-0003-3460-0103]{Alexei V. Filippenko}
\affiliation{Department of Astronomy, University of California, Berkeley, CA 94720-3411, USA }
\affiliation{Miller Institute for Basic Research in Science, University of California, Berkeley, CA 94720, USA}

\author[0000-0001-5975-290X]{Joel Johansson}
\affiliation{Oskar Klein Centre, Department of Physics, Stockholm University, SE 106 91 Stockholm, Sweden}

\author{Patrick L. Kelly}
\affiliation{School of Physics and Astronomy, University of Minnesota, 116 Church Street SE, Minneapolis, MN 55455, USA}

\author{Kelsie Krafton}
\affiliation{Department of Physics \& Astronomy, Louisiana State University, Baton Rouge, LA 70803, USA}

\author[0000-0001-5788-5258]{A. P. Marston}
\affiliation{ESA/JWST, Space Telescope Science Institute, 3700 San Martin Drive, Baltimore, MD 21218, USA}

\author[0000-0002-7555-8741]{Jon C. Mauerhan}
\affiliation{The Aerospace Corporation, 2310 E. El Segundo Blvd., El Segundo, CA 90245, USA}

\author[0000-0001-9038-9950]{Schuyler D.~Van Dyk}
\affiliation{Caltech/IPAC, Mailcode 100-22, 1200 E.~California Blvd., Pasadena, CA 91125, USA}

\begin{abstract}

Here we present new -- and, nevertheless, last -- mid-infrared (mid-IR) data for supernovae (SNe) based on measurements with the {\it Spitzer Space Telescope}. Comparing our recent 3.6 and 4.5\,$\mu$m photometry with previously published mid-IR and further multiwavelength datasets, we were able to draw some conclusions about the origin and heating mechanism of the dust in these SNe or in their environments, as well as on possible connection with circumstellar matter (CSM) originating from pre-explosion mass-loss events in the progenitor stars.
We also present new results regarding both certain SN classes and single objects. We highlight the mid-IR homogeneity of SNe Ia-CSM, which may be a hint of their common progenitor type and of their basically uniform circumstellar environments. Regarding single objects, it is worth highlighting the late-time interacting Type Ib SNe 2003gk and 2004dk, for which we present the first-ever mid-IR data, which seem to be consistent with clues of ongoing CSM interaction detected in other wavelength ranges.
Our current study suggests that long-term mid-IR follow-up observations play a key role in a better understanding of both pre- and post-explosion processes in SNe and their environments. While {\it Spitzer} is not available any more, the expected unique data from the {\it James Webb Space Telescope}, as well as long-term near-IR follow-up observations of dusty SNe, can bring us closer to the hidden details of this topic.

\end{abstract}


\section{Introduction} \label{sec:intro}


Supernovae (SNe) are the final explosions of either evolved massive stars or of white dwarfs (WDs) located in binary systems.  These events are important astrophysical laboratories for studying not just the details of the cataclysmic endings of stars, but also the preceding stellar evolution processes and the effect of SNe on their surrounding environments. Either early-time (so-called ``flash") spectroscopy, or long-term (multichannel) follow-up observations of the SN ejecta and their interaction with the surrounding circumstellar medium (CSM) can provide important details about the pre-explosion mass-loss history of the progenitors and the physics of shock waves. Beyond that, late-time observations can allow us to reveal the signs of special astrophysical processes may take place during the interaction (e.g., ionization and recombination of the gas, or heating/formation of dust grains).

Both the timescale and degree of ejecta-CSM interaction vary among explosions and, specifically, SN type and subclass. Type IIn SNe were the original class of SNe exhibiting clear signatures of interaction between the ejecta and dense, nearby CSM (e.g., relatively narrow optical emission lines denoted by ``$n$" in SNe~IIn, as well as strong X-ray and radio emission) within days after explosion \citep{schlegel90}. While the origin of SNe~IIn is generally explained with core-collapse (CC) explosions of massive stars into previously expelled H-rich environments, similar phenomena can happen in the presence of a dense He-rich CSM (called SNe~Ibn). Additionally, progressively more ``normal" CC or thermonuclear SNe (i.e., exploding WDs) are found to eventually start producing detectable signs of interaction (occurring in CSM shells at larger distances from the explosion site); see, e.g., \citet{vinko17}, as well as the recent reviews of \citet{cf17} and \citet{smith_17hsn}.

Existing ground-based transient surveys ensure the optical monitoring of hundreds of SNe every year; however, these observations focus mostly on the ``early'' ($\lesssim 3$\,yr) phases of the objects. Late-time observations (either optical or at other wavelengths) are mostly occasional because they require large-aperture or space telescopes. Mid-infrared (mid-IR) observations at late times, in particular, offer several advantages over optical SN observations: increased sensitivity to the ejecta as they expand and cool, less impact from interstellar extinction, and coverage of atomic and molecular emission lines generated by shocked, cooling gas \citep[see, e.g.,][]{reach06}. Perhaps most important, mid-IR observations are also sensitive to warm dust in the SN environment.

The origin and heating mechanism of the dust, however, is not always obvious, as the dust may be newly formed or pre-existing in the CSM. Newly-condensed dust may form in either the ejecta or in a cool dense shell (CDS) between the shocked CSM and shocked ejecta where material cools \citep[see, e.g.,][]{pozzo04,mattila08}. Pre-existing dust may be radiatively heated by the peak SN luminosity or by the radiation from the shock breakout \citep[e.g.,][]{dwek08}, or by energetic photons generated during late-time CSM interaction, thereby forming an IR echo \citep[see, e.g.,][]{bode80,dwek83,graham86,sugerman03,kotak09}. In this case, the dust is a useful probe of the CSM characteristics and the pre-SN mass loss from either the progenitor or companion star \citep[see, e.g.,][for a review]{gall11}.



In the last 1.5 decades, the {\it Spitzer Space Telescope} (hereafter {\it Spitzer}) -- especially its InfraRed Array Camera (IRAC) detector \citep{fazio04} -- was the primary source of mid-IR SN observations. Between 2003 and 2009, during the cryogenic (or Cold Mission) phase, only a moderate number ($<50$) of nearby SNe were targeted by {\it Spitzer}. After 2009, by this time with post-cryo (Warm Mission) {\it Spitzer} (with the availability of the two shortest-wavelength IRAC channels at 3.6 and 4.5\,$\mu$m), two large surveys contributed to this surge: a program aimed to observe a large sample of Type IIn SNe \citep[$\sim70$ observed SN sites, 13 detected targets; see][]{fox11,fox13}, and SPitzer InfraRed Intensive Transients Survey (SPIRITS), a systematic mid-IR study of nearby ($\lesssim20$\,Mpc) galaxies. SPIRITS has resulted in the detection of $\sim50$ SNe of various types \citep[][]{tinyanont16,kasliwal17,jencson19}, including obscured SNe missed by previous optical surveys \citep{jencson17,jencson18,jencson19}, as well as some other transients showing unusual IR behavior \citep{kasliwal17}.
Beyond these studies and a number of single-object papers, further broader studies were also presented regarding the mid-IR behavior of SNe~II-P \citep{szalai13} and SNe~Ia \citep{johansson17}.

Most recently, some of us \citep{szalai19} published a comprehensive analysis of the largest mid-IR dataset of SNe ever studied ({\it Spitzer} data for $\sim 1100$ SN sites including 120 positive detections). This sample -- including all previously published data on most SNe discovered up to 2015, as well as many further objects that appeared on nontargeted {\it Spitzer} images -- allowed an in-depth analysis to constrain the origin and heating mechanism of the dust in each SN type, and to perform preliminary statistics on their long-term mid-IR evolution.
One of the main findings of this study was that subtypes of CC~SNe tend to fill their own regions
of phase space (IR luminosity, dust temperature, dust mass), while in thermonuclear SNe, there is a huge gap in late-time mid-IR properties of the few strongly interacting SNe and the only slightly detected or nondetected objects. Furthermore, SNe~IIn and other strongly interacting SNe remain bright for several years after explosion in the mid-IR, probably due to radiation from a large amount ($\gtrsim 10^{-3}\,M_{\odot}$) of pre-existing, radiatively heated dust grains.

Up to its decommissioning in January 2020, {\it Spitzer} has continued observations of SN sites in the framework of the SPIRITS project \citep[see the latest summary of its results in][]{jencson19}, or via following nearby, high-priority targets like the dust-forming Type II-P SN~2017eaw \citep{tinyanont19b,szalai19b}.
A targeted SN survey by our group, LASTCHANCE, was also carried out between August 2018 and January 2020, aiming to collect further mid-IR photometric data points on a number of SNe in order to examine the long-term evolution of different types of stellar explosions.


In this paper, we present the results of our survey, focusing primarily on interacting SNe which constitute the vast majority of the targets we observed.
In Section \ref{sec:obs}, we describe the steps of the data collection and photometry of {\it Spitzer}/IRAC data. We present our results in Section~\ref{sec:res}, and our conclusions are discussed in Section~\ref{sec:concl}.

\section{Observations \& Data Analysis} \label{sec:obs}

\subsection{Targets of Our Study}
\label{sec:obs_targ}

During our {\it Spitzer}/LASTCHANCE program (PID 14098; PI O.~D. Fox), we observed 31 targets meeting one of the following criteria: (i) high-profile, well-understood interacting and/or dust-forming SNe that were previously detected by {\it Spitzer}/IRAC and required another epoch to monitor their mid-IR evolution \citep[based on][and references therein]{szalai19}, or (ii) young ($\lesssim 1$\,yr) and relatively nearby ($\lesssim 70$\,Mpc) SNe with known dense CSM (from optical spectra) or with a potential of dust formation in their ejecta (Type II-P SNe). 


\begin{figure*}
\begin{center}
\includegraphics[width=.8\textwidth]{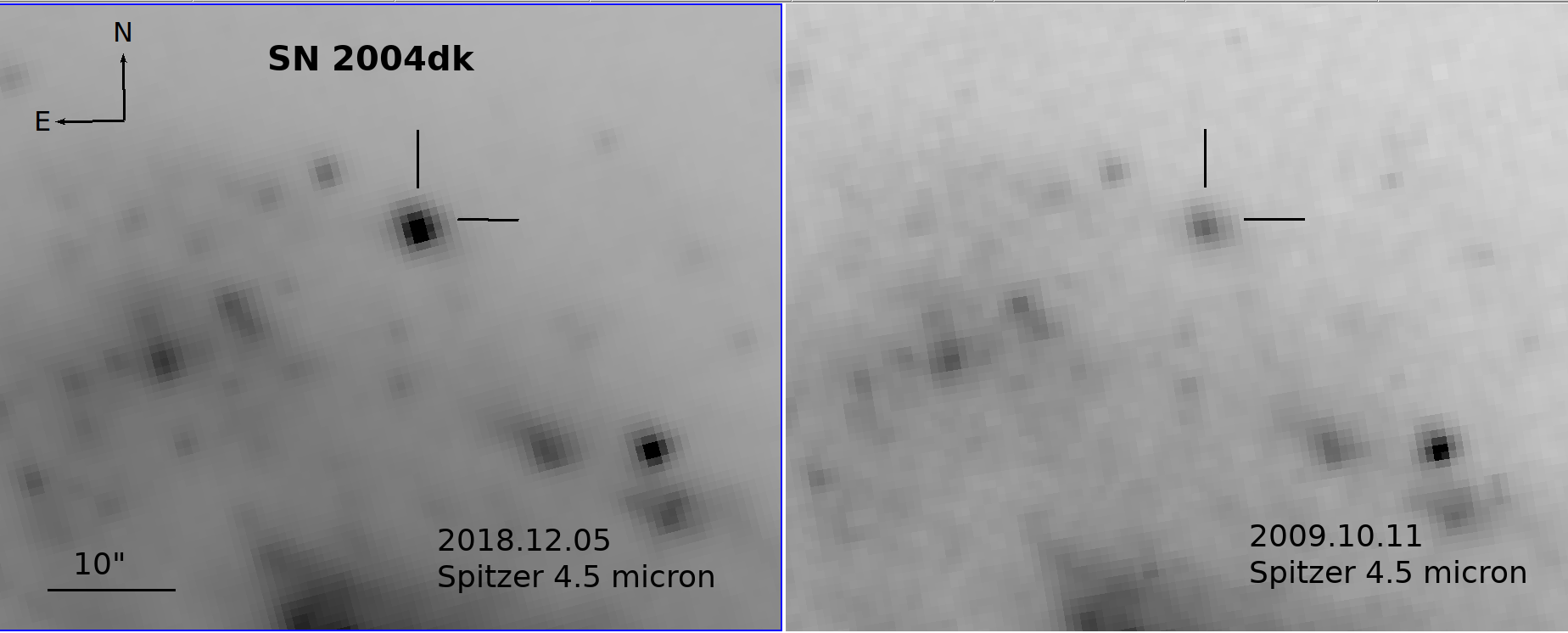}
\includegraphics[width=.8\textwidth]{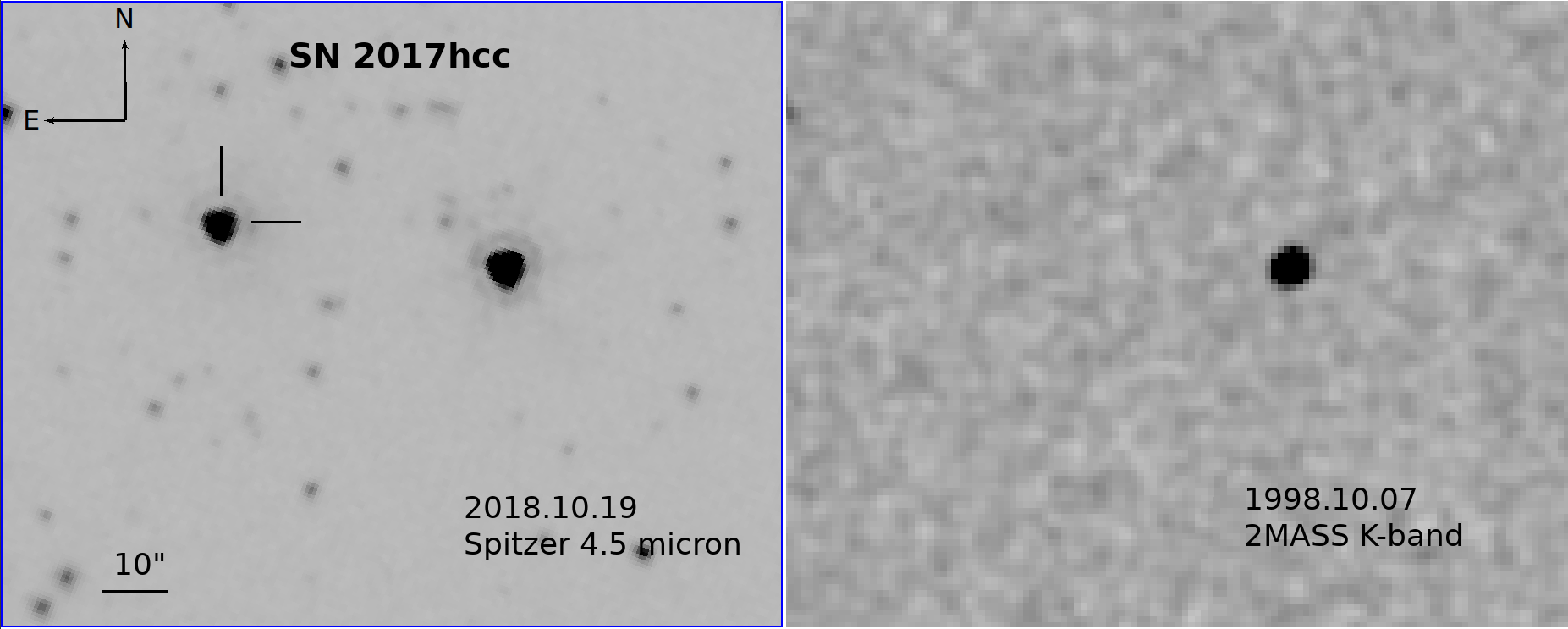}
\includegraphics[width=.8\textwidth]{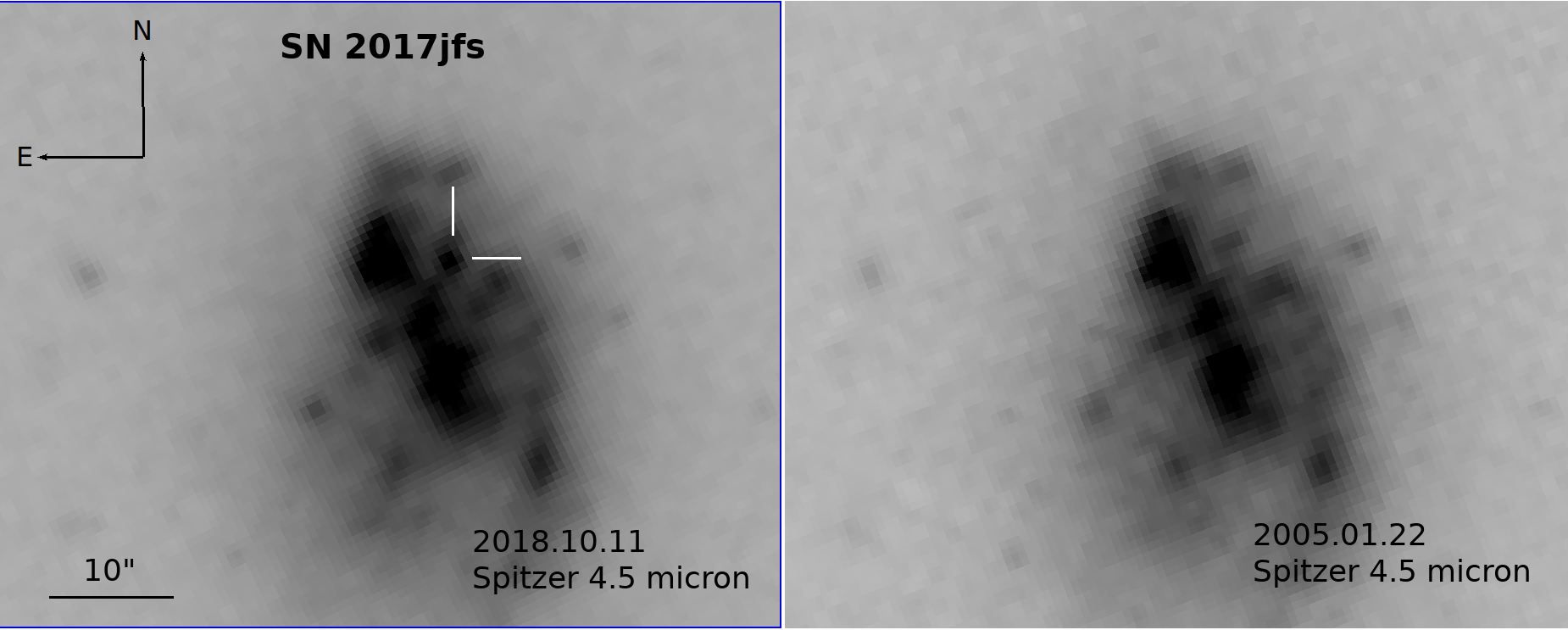}
\caption{Examples of new positive detections in {\it Spitzer}/IRAC 4.5\,$\mu$m images. {\it Top:} SN 2004dk (Type Ib, post-explosion images after {\it (left)} and before {\it (right)} mid-IR rebrightening). {\it Middle:} SN 2017hcc (Type IIn, post-explosion image {\it (left)} and archival 2MASS $K_s$-band image ({\it right})). {\it Bottom}: SN 2017jfs (LRN, post-explosion {\it (left)} and pre-explosion {\it (right)} {\it Spitzer} images). 
}
\label{fig:irac}
\end{center}
\end{figure*}

\begin{deluxetable*}{lcccccccc}
\tabletypesize{\scriptsize}
\tablecaption{\label{tab:sn} Basic data of the studied SNe.}
\tablehead{
\colhead{Object} & \colhead{SN Type} & \colhead{Discovery} & \colhead{Host} & \colhead{$\alpha$ (J2000)} & \colhead{$\delta$ (J2000)} & \colhead{$d$} & \colhead{$E(B-V)^{\dagger}$} & \colhead{Refe-} \\
\colhead{} & \colhead{} & \colhead{(MJD)} & \colhead{Galaxy} & \colhead{} & \colhead{} & \colhead{(Mpc)} & \colhead{(mag)} & \colhead{rences}
}
\startdata
SN~1995N & IIn & 49842 & MCG-02-38-17 & 14:49:28.31 & $-$10:10:14.0 & 24.0 $\pm$ 4.0 & 0.10 & 1 \\
SN~2001em & Ib/c & 52172 & UGC 11794 & 21:42:23.60 & $+$12:29:50.3 & 71.6 $\pm$ 0.7 & 0.10 & 2, 3 \\
SN~2003gk & Ib & 52821 & NGC 7460 & 23:01:42.99 & +02:16:08.7 & 41.2 $\pm$ 6.0 & 0.08 & 4 \\
SN~2004dk & Ib & 53218 &  NGC 6118 & 16:21:48.93 & $-$02:16:17.3 & 24.0 $\pm$ 2.0 & 0.14 & 5 \\
SN~2005ip & IIn & 53679 & NGC 2906 & 09:32:06.42 & +8:26:44.4 & 30.0 $\pm$ 7.2 & 0.04 & 6 \\
SN~2006jd & IIn & 54020 & UGC 4179 & 08:02:07.43 & $+$00:48:31.5 & 77.0 $\pm$ 5.0 & 0.05 & 7 \\
SN~2009ip & IIn / imp. & 56193 & NGC 7259 & 22:23:08.30 & $-$28:56:52.0 & 20.5 $\pm$ 2.0 & 0.02 & 7 \\
SN~2010jl & IIn & 55503 & UGC 5189A & 09:42:53.33 & $+$09:29:41.8 & 48.9 $\pm$ 3.4 & 0.02 & 8 \\
SN~2010mc & IIn & 55428 & anon. & 17:21:30.68 & $+$48:07:47.4 & 159.0$^{\ddagger}$ & 0.02 & 9, 10 \\
SN~2011ft & Ib & 55803 & UGC 11021 & 17:52:42.98 & $+$29:04:10.6  & 101.0 $\pm$ 3.0 & 0.06 & 11, 12 \\
PTF11iqb & IIn & 55765 & NGC 151 & 00:34:04.84 & $-$09:42:17.9 & 42.3 $\pm$ 11.5 & 0.03 & 13 \\
PTF11kx & Ia-CSM & 55579 & anon. & 08:09:12.87 & $+$46:18:48.8 & 200.0$^{\ddagger}$ & 0.04 & 14 \\
SN 2012aw & II-P & 56002 & NGC 3351 & 10:43:53.73 & $+$11:40:17.6 & 8.8 $\pm$ 1.1 & 0.02 & 15 \\
SN~2012ca & Ia-CSM & 56042 & ESO 336-09 & 18:41:07.25 & $-$41:47:38.4 & 80.0 $\pm$ 6.0$^{\ddagger}$ & 0.07 & 16, 17 \\
SN~2013L & IIn & 56314 &  ESO 216-G39 & 11:45:29.55 & $-$50:35:53.1 & 75.0 $\pm$ 5.0$^{\ddagger}$ & 0.11 & 18 \\
SN~2013cj & IIn & 56421 & UGC 10685 & 17:04:52.95 & $+$12:55:10.4 & 135.0 $\pm$ 10.0$^{\ddagger}$ & 0.09 & 19 \\ 
SN 2013ej & II-P/L & 56497 & NGC 628 & 01:36:48.16 & $+$15:45:31.0 & 9.5 $\pm$ 0.6 & 0.06 & 20 \\
SNHunt248 & IIn / imp. & 56798 & NGC 5806 & 14:59:59.47 & +01:54:26.6 & 20.0 $\pm$ 3.0$^{\ddagger}$ & 0.04 & 21 \\
ASSASN-14dc	& IIn & 56832 & PGC 2035709 & 02:18:37.82 & +33:37:01.7 & 183.0 $\pm$ 20.0$^{\ddagger}$ & 0.06 & 22 \\
SN~2015da & IIn & 57031 & NGC 5337 & 13:52:24.11 & +39:41:28.6 & 30.0 $\pm$ 10.0$^{\ddagger}$ & 0.02 & 23 \\
AT2016jbu & IIn / imp. & 57723 & NGC 2442 & 07:36:25.96 & $-$69:32:55.3 & 21.0 $\pm$ 1.5 & 0.2 & 24, 25 \\
SN~2017aym & II-P & 57766 & NGC 5690 & 14:37:41.78 & +02:17:08.4 & 18.6 $\pm$ 3.0 & 0.04 & 3, 26 \\
SN~2017eaw & II-P & 57887 & NGC 6946 & 20:34:44.24 & +60:11:36.0 & 6.9 $\pm$ 0.6 & 0.4 & 27 \\
SN~2017ejx & II-P & 57903 & NGC 2993 & 09:45:48.61 & $-$14:22:05.7 & 30.5 $\pm$ 5.0$^{\ddagger}$ & 0.05 & 28, 29 \\
SN~2017gas & IIn & 57975 & anon. & 20:17:11.32 & +58:12:08.0 & 42.0 $\pm$ 5.0$^{\ddagger}$ & 0.33 & 30 \\
SN~2017hcc & IIn & 58028 & anon. & 00:03:50.58 & $-$11:28:28.8 & 72.0 $\pm$ 6.0$^{\ddagger}$ & 0.03 & 31 \\
SN~2017ivu & II-P & 58098 & NGC 5962 & 15:36:32.70 & +16:36:19.4 & 30.7 $\pm$ 4.0 & 0.05 & 3, 32 \\
SN~2017jfs & IIn / LRN & 58113 & NGC 4470 & 12:29:37.79 & +07:49:35.2 & 33.0 $\pm$ 5.0 & 0.02 & 33, 34 \\
SN~2018gj & II & 58132 & NGC 6217 & 16:32:02.30 & +78:12:40.9 & 24.0 $\pm$ 3.0 & 0.04$^{\ddagger}$ & 35 \\
SN~2018zd & IIn & 58179 & NGC 2146 & 06:18:03.18 & +78:22:00.9 & 15.2 $ \pm$ 4.0 & 0.08 & 36, 37 \\
SN~2018acj & II-P & 58185 & UGC 8733 & 13:48:40.63 & +43:25:04.7 & 32.1 $\pm$ 5.0 & 0.02 & 3, 38 \\
SN~2018fhw & Ia-CSM(?) & 58351 & anon. & 04:18:06.20 &	$-$63:36:56.4 & 74.2 $\pm$ 4.0 & 0.03 & 39 \\
\enddata
\tablecomments{{\it References}:
$^1$\citet{VD13}; $^2$\citet{Papenkova01}; $^3$\citet{sorce14}; 
$^4$\citet{theureau07};
$^5$\citet{springob09}; $^6$\citet{fox10};
$^7$\citet{fox11}; 
$^8$\citet{fox13}; $9$\citet{howell12}; $^{10}$\citet{ofek13}; $^{11}$\citet{balanutsa11}; $^{12}$\citet{prieto11}; $^{13}$\citet{parrent11}; $^{14}$\citet{graham17};  $^{15}$\citet{siviero12};
$^{16}$\citet{drescher12}; $^{17}$\citet{inserra12}; $^{18}$\citet{tinyanont16}; $^{19}$\citet{jin13}; $^{20}$\citet{mauerhan17};
$^{21}$\citet{mauerhan15}; $^{22}$\citet{holoien14}; $^{23}$\citet{zhang15}; $^{24}$\citet{cartier17}; $^{25}$\citet{fraser17}; $^{26}$\citet{taddia17} -- ATel 10148; $^{27}$\citet{szalai19b}; $^{28}$\citet{brimacombe17}; $^{29}$\citet{KR17}; $^{30}$\citet{balam17}; $^{31}$\citet{dong17}; $^{32}$\citet{itagaki17}; $^{33}$\citet{berton18}; $^{34}$\citet{pastorello19a}; $^{35}$\citet{kilpatrick18}; $^{36}$\citet{zhang18}; $^{37}$\citet{gao04}; $^{38}$\citet{lin18};
$^{39}$\citet{vallely19}.
{\it Additional notes}:$^{\dagger}$ Galactic extinction. $^{\ddagger}$ Distance calculated from redshift.
}
\end{deluxetable*}


Our observations were designed to attain a signal-to-noise ratio (S/N) of 100 for a medium background according to the {\it Spitzer}/IRAC Sensitivity Performance Estimation Tool (SENS-PET). We aimed for this S/N so that any photometric uncertainties would be dominated by systematic effects associated with the background (host-galaxy) subtraction rather than insufficient integration time. This S/N also provides a buffer for detecting SNe that may have faded more quickly than predicted. Our observations used the medium-scale cycling dither pattern to provide good redundancy and capability for self-calibration.

Basic data on the SNe and their hosts -- collected via the Open Supernova Catalog\footnote{\href{https://sne.space}{https://sne.space}} \citep{guillochon17}, Simbad database\footnote{\href{http://simbad.u-strasbg.fr/simbad/}{http://simbad.u-strasbg.fr/simbad/}} \citep{wenger00}, and NASA/IPAC Extragalactic Database\footnote{\href{http://ned.ipac.caltech.edu/}{http://ned.ipac.caltech.edu/}} -- are shown in Table \ref{tab:sn}. All of our LASTCHANCE targets were observed at one epoch -- except SNe 2004dk, 2018fhw (2 epochs), and 2017hcc (3 epochs) -- in both the 3.6 and 4.5\,$\mu$m channels.
We also checked the {\it Spitzer} Heritage Archive (SHA)\footnote{\href{http://sha.ipac.caltech.edu}{http://sha.ipac.caltech.edu}} for pre-explosion {\it Spitzer}/IRAC images of the SN sites, to be used as templates (see next section). 

 

\subsection{Object Identification and Photometry}\label{obs_phot}

We collected and analyzed IRAC post-basic calibrated data (PBCD). The scale of these images is 0.6$\arcsec$\,pixel$^{-1}$. 
Photometric analysis was carried out using the \texttt{phot} task of IRAF\footnote{IRAF is distributed by the National Optical Astronomy Observatories, which are operated by the Association of Universities for Research in Astronomy, Inc., under cooperative agreement with the National Science Foundation (NSF).}. For isolated sources, we implemented aperture photometry on the PBCD frames using the \texttt{phot} task as a first step. We generally used an aperture radius of 2\arcsec\ and a background annulus from 2\arcsec\ to 6\arcsec\ (2$-$2$-$6 configuration), and applied aperture corrections of 1.213, 1.234, 1.379, and 1.584 for the four IRAC channels (3.6, 4.5, 5.8, and 8.0\,$\mu$m, respectively) as given in the IRAC Data Handbook; however, for a few bright sources extending to more pixels, we used the 3$-$3$-$7 configuration (aperture corrections: 1.124, 1.127, 1.143, and 1.234, respectively). 


Identifying a point source within a host galaxy, where compact \ion{H}{2} regions and star clusters
may also appear as point-like sources in the images, can be difficult, especially if the target is faint or is superposed on a complex background. Therefore, when available, we also performed aperture photometry on template images (i.e., pre-explosion images, or very late-time images in which no point source appears at the position of the SN) and subtracted these on-site flux values from the SN fluxes. We call this technique ``template-based background subtraction," or simply ``background subtraction" hereafter.

Not all targets have templates. In these cases, the local background was estimated by measuring the nearby flux with apertures placed adjacent to the SN site in locations that qualitatively appear similar to the background underlying the SN \citep[as was applied, e.g., by][]{fox11}.  More complex modeling of the local background is beyond the scope of this paper.

In both background subtracted and not subtracted cases, we defined the source as a positive detection if (i) it showed epoch-to-epoch flux changes, and (ii) its flux was {\it above} the local background by at least 5\,$\mu$Jy and 15\,$\mu$Jy at 3.6 and 4.5\,$\mu$m, respectively (according to point-source sensitivities in Table 2.10 of the IRAC Instrument Handbook\footnote{\href{https://irsa.ipac.caltech.edu/data/SPITZER/docs/irac/iracinstrumenthandbook/}{https://irsa.ipac.caltech.edu/data/SPITZER/docs/irac/\\iracinstrumenthandbook/}}).

Following these steps, we find that 19 of 31 targets are detected (one object, SN~1995N, only at 4.5 $\mu$m). In three other cases (SNe~2011ft, 2017gas, and 2018fhw), there is a point source close to the coordinates of the SN; however, given the lack of template images, they cannot be confirmed. SN~2017gas is very close to the core, while SN~2011ft is a distant object that cannot be distinguished from its host \citep[note that this latter site was also captured $\sim 6$\,yr earlier with {\it Spitzer}, but only at 3.6\,$\mu$m, and there are no significant flux changes compared to this previously published value; see][]{szalai19}. We discuss the case of SN~2018fhw later.

\begin{deluxetable*}{lccrcccc}
\tablecaption{Previously unreported mid-IR ({\it Spitzer}) detections and nondetections (fluxes and Vega magnitudes) of interacting SNe. We show here all new measurements from our LASTCHANCE program, together with unpublished archive data of SNe 2003gk and 2017eaw. \label{tab:phot}}
\tabletypesize{\scriptsize}
\tablehead{
\colhead{Object} & \colhead{Type} & \colhead{Date} & \colhead{Epoch$^{\dagger}$} & \colhead{$F_{\nu,[3.6]}$} & \colhead{$F_{\nu,[4.5]}$} & \multicolumn{2}{c}{Absolute mag.} \\
\colhead{} & \colhead{} & \colhead{(MJD)} & \colhead{(days)} & \colhead{($\mu$Jy)} & \colhead{($\mu$Jy)} & \colhead{3.6 $\mu$m} & \colhead{4.5 $\mu$m}
}
\startdata
{\it SN~1995N} & {\it IIn} & {\it 58419} & {\it 8577} & {\it $<5$} & {\it 13(5)} & {\it $-$12.5$<$} & {\it $-$14.06(0.62)} \\
SN~2001em$^a$ & Ib/c & 53307 & 1135 & 221(41) & 294(36) & $-$19.02(0.20) & $-$19.81(0.13) \\
~ & ~ & {\it 58541}	& {\it 6369} & {\it $<$82} & {\it $<$55} & {\it $-$17.9$<$} & {\it $-$18.0$<$} \\
{\it SN~2003gk$^b$} & {\it Ib} & {\it 55939} & {\it 3118} & ... & {\it 2438(83)} & ... & {\it $-$20.45(0.31)} \\
~ & ~ & {\it 56160} & {\it 3339} & ... & {\it 2062(75)} & ... & {\it $-$20.27(0.31)} \\
~ & ~ & {\it 56536} & {\it 3715} & ... & {\it 1510(65)} & ... & {\it $-$19.93(0.32)} \\
~ & ~ & {\it 56699} & {\it 3878} & ... & {\it 1336(61)} & ... & {\it $-$19.79(0.32)} \\
~ & ~ & {\it 56912} & {\it 4091} & ... & {\it 1123(56)} & ... & {\it $-$19.61(0.32)} \\
~ & ~ & {\it 57819} & {\it 4998} & ... & {\it 651(43)} & ... & {\it $-$19.02(0.33)} \\
~ & ~ & {\it 58031} & {\it 5210} & ... & {\it 592(42)} & ... & {\it $-$18.91(0.33)} \\
~ & ~ & {\it 58204} & {\it 5383} & ... & {\it 556(39)} & ... & {\it $-$18.84(0.33)} \\
{\it SN~2004dk} & {\it Ib} & {\it 55115} & {\it 1897} & {\it 85(16)} & {\it 93(16)} & {\it $-$15.62(0.28)} & {\it $-$16.19(0.26)} \\
~ & ~ & {\it 58457} & {\it 5239} & {\it 141(20)} & {\it 201(24)} & {\it $-$16.17(0.24)} & {\it $-$17.03(0.22)} \\
~ & ~ & {\it 58852} & {\it 5634}& {\it 130(20)} & {\it 183(23)} & {\it $-$16.08(0.24)} & {\it $-$16.93(0.22)} \\
{\it SN~2005ip} & {\it IIn} & {\it 58346} & {\it 4667} & {\it 677(51)} & {\it 713(47)} & {\it $-$18.34(0.52)} & {\it $-$18.88(0.52)} \\
{\it SN~2006jd} & {\it IIn} & {\it 58345} & {\it 4325} & {\it 82(15)} & {\it 106(17)} & {\it $-$18.11(0.24)} & {\it $-$18.86(0.22)} \\
{\it SN~2009ip} & {\it IIn / imp.} & {\it 58529} & {\it 2336} & {\it $<$3} & {\it $<$5} & {\it $-$11.6$<$} & {\it $-$12.7$<$} \\
SN~2010jl & IIn & 58372 & 2869 & 300(42) & 611(47) & $-$18.52(0.21) & $-$19.78(0.17) \\
{\it SN~2010mc} & {\it IIn} & {\it 58411} & {\it 2983} & {\it 33(9)} & {\it 37(10)} & {\it $-$18.72(0.34)} & {\it $-$19.31(0.32)} \\
{\it PTF11iqb} & {\it IIn} & {\it 57089} & {\it 1324} & {\it 120(20)} & {\it 143(21)} & {\it $-$17.21(0.61)} & {\it $-$17.89(0.61)} \\
~ & ~ &  {\it 58405} & {\it 2640} & {\it 64(15)} & {\it 51(13)} & {\it $-$16.54(0.64)} & {\it $-$16.76(0.65)} \\
{\it PTF11kx} & {\it Ia-CSM} & {\it 58516} & {\it 2866} & {\it $<$5} & {\it $<$15} & {\it $-$17.1$<$} & {\it $-$18.8$<$} \\
SN~2012aw & II-P & 58366 & 2364 & $<5$ & $<15$ & $-10.4<$ & $-12.0<$ \\
{\it SN~2013L} & {\it IIn} & {\it 58405} & {\it 2091} & {\it 289(28)} & {\it 506(37)} & {\it $-$19.42(0.18)} & {\it $-$20.50(0.16)} \\
{\it SN~2012ca} & {\it Ia-CSM} & {\it 58490} & {\it 2448} & {\it $<$5} & {\it $<$15} & {\it $-$15.2$<$} & {\it $-$16.8$<$} \\
{\it SN~2013cj} & {\it IIn} & {\it 58457} & {\it 2036} & {\it 119(18)} & {\it 214(24)} & {\it $-$19.73(0.23)} & {\it $-$20.84(0.20)} \\
SN~2013ej & II-P/L & 58427 & 1930 & $<5$ & $<15$ & $-10.5<$ & $-12.2<$ \\
SNHunt248 & IIn / imp. & 58419 & 1621 & $<5$ & $<15$ & $-12.1<$ & $-13.8<$ \\
ASSASN-14dc$^c$	& IIn & 57323 & 491 & 176(37) & 217(35) & $-$20.81(0.33) & $-$21.51(0.29) \\
~ & ~ & {\it 58466} & {\it 1634} & {\it $<$320} & {\it $<$220} & {\it $-$20.7$<$} & {\it $-$20.7$<$} \\
SN~2015da & IIn & 58404 & 1373 & 1603(69) & 2155(79) & $-$19.28(0.72) & $-$20.08(0.72) \\
AT2016jbu & IIn/imp. & 58352 & 629 & 67(16) & 90(18) & $-$15.07(0.31) & $-$15.87(0.27) \\
SN~2017aym & II-P & 58415 & 649 & 265(79) & 303(59) & $-$16.29(0.47) & $-$16.91(0.41) \\
SN~2017eaw & II-P & 58404 & 517 & 124(18) & 314(30) & $-$13.07(0.25) & $-$14.76(0.21) \\
~ & ~ & 58488$^d$ & 601 & 78(15) & 149(20) & $-$12.83(0.25) & $-$14.00(0.22) \\
~ & ~ & 58515$^e$ & 628 & 79(15) & 128(20) & $-$12.85(0.28) & $-$13.83(0.22) \\
~ & ~ & 58529$^d$ & 642 & 75(16) & 132(19) & $-$12.79(0.28) & $-$13.87(0.24) \\
~ & ~ & 58567$^d$ & 680 & 58(14) & 83(18) & $-$12.51(0.28) & $-$13.37(0.24) \\
~ & ~ & 58607$^d$ & 720 & 88(16) & 121(19) & $-$12.96(0.27) & $-$13.78(0.25) \\
~ & ~ & 58795$^d$ & 908 & 49(12) & 57(13) & $-$12.34(0.32) & $-$12.97(0.31) \\
SN~2017ejx & II-P/L & 58373 & 470 & $<5$ & $<15$ & $-13.1<$ & $-14.7<$ \\
{\it SN~2017hcc} & {\it IIn} & {\it 58420} & {\it 392} & {\it 1818(70)} & {\it 1813(71)} & {\it $-$21.32(0.18)} & {\it $-$21.79(0.18)} \\
~ & ~ & {\it 58593} & {\it 565}  & {\it 1533(64)} & {\it 1703(69)} & {\it $-$21.13(0.18)} & {\it $-$21.72(0.18)} \\
~ & ~ & {\it 58795} & {\it 767}  & {\it 1097(55)} & {\it 1354(62)} & {\it $-$20.77(0.18)} & {\it $-$21.47(0.18)} \\
SN~2017ivu & II-P & 58416 & 318 & 41(15) & 68(20) & $-$15.35(0.37) & $-$16.38(0.31) \\
SN~2017jfs & IIn/LRN & 58402 & 289 & 100(44) & 63(31) & $-$16.48(0.58) & $-$16.46(0.63) \\
SN~2018gj & II & 58484 & 352 & 17(7) & 83(15) & $-$13.91(0.50) & $-$16.06(0.33) \\
SN~2018zd & IIn & 58507 & 328 & 40(13) & 73(15) & $-$13.80(0.68) & $-$14.93(0.62) \\
SN~2018acj & II-P & 58391 & 206 & 15(6) & 27(8) & $-$14.41(0.57) & $-$15.50(0.48) \\
\enddata
\tablecomments{Data marked with italics denote cases where template-based background subtraction -- given the lack of either pre-explosion or late-time images -- cannot be applied (in all these cases, measured fluxes can be considered only as upper limits).
$^{\dagger}$ Days since discovery. {\it Additional comments}: $^a$Corrected fluxes/magnitudes determined after background subtraction based on recently obtained {\it Spitzer} images in which the target is not detectable; data paper contains the original fluxes: \citet{szalai19}. $^b$Target detection (SN~2003gk) in archive images: PIDs 90007 and 13012 (PI J.~D. Kirkpatrick). $^c$Positive detections in archive images after background subtraction based on recently obtained {\it Spitzer} images in which the target is not detectable: ASASSN-14dc -- PID 11053 (PI O. Fox). 
$^d$Archive data (PID 14089, PI M. Kasliwal). $^e$Archive data (PID 13239, PI K. Krafton).
}
\end{deluxetable*}

In the cases of the remaining 9 targets, no point sources are detected at the position of the SN. Instead, some of these nondetections provided us with template images, which allowed us to carry out background subtraction on earlier archive data of these particular targets. 
In the cases of PTF11kx, SNHunt248, ASASSN-14dc, SN 2009ip, and SN 2012ca, background subtractions do not change the previously published flux values \citep[see][and references therein]{szalai19} beyond photometric uncertainties. 
Instead, the change in measured flux is more significant in the case of SN 2001em \citep[221$\pm$41 and 294$\pm$36 $\mu$Jy at 3.6 and 4.5 $\mu$m, respectively, vs. 303$\pm$41 and 349$\pm$37 $\mu$Jy given in][]{szalai19} located in a distant, nearly edge-on galaxy. We included updated fluxes and correponding absolute magnitudes of SN~2001em in Table \ref{tab:phot}.

Moreover, during the review of archival {\it Spitzer} data, we identified a variable mid-IR source at the position of the Type Ib SN~2003gk. While there are no published {\it Spitzer} data for this object \citep[we missed it during our overview in][]{szalai19}, we added it to our sample and analyzed it the same way as the other targets (discussed in more detail below). In the case of the Type II-P SN~2017eaw, we also present here the previously unpublished photometry of late-time archival data. 

In summary, there are 32 objects we studied during this program: 13 SNe IIn (11 positive detections, 1 unconfirmed detection, 1 nondetection), 7 SNe II-P (4 positive detections, 3 nondetections), 1 unclassified SN II (positive detection), 4 SNe Ib/c (2 positive detections, 1 unconfirmed detection, 1 nondetection), 3 SNe Ia-CSM (1 unconfirmed detection, 2 nondetections), and 4 intermediate-luminosity transients (2 positive detections, 2 nondetections). These detection rates -- taking into account epochs and distances -- are basically in agreement with the expectations based on earlier results \citep{tinyanont16,szalai19}; see details later.
For 12 of our targets, no {\it Spitzer} data have been published before (Type IIn PTF11iqb, ASASSN-14dc, SN~2015da, and SN~2017hcc; Type II-P SNe 2017aym, 2017ivu, 2018gj, 2018acj; SNe Ib 2003gk and 2004dk; and intermediate-luminosity transients AT2016jbu and SN~2017jfs).

In Figure~\ref{fig:irac}, we present examples of new positive detections (SNe 2004dk, 2017hcc, and 2017jfs).
Mid-IR fluxes and cases of nondetections are listed in Table~\ref{tab:phot}. We also identify SNe where background subtraction was not possible.  We note that in all these cases, the measured fluxes should be interpreted as upper limits given the complex underlying backgrounds. Flux uncertainties are generally based on photon statistics provided by \texttt{phot}; however, where background-subtraction photometry was carried out, an increase of the noise level by $\sqrt{2}$ is also taken into account.
%

\section{Results} \label{sec:res}

\citet{szalai19} first published the comprehensive mid-IR light curves (LCs) of SNe based on their complete datasets available at that time.  In this work, we present new {\it Spitzer} data points for several SNe.  Furthermore, we compare the long-term mid-IR evolution of different types of interacting SNe. We also present our findings on connections between mid-IR behavior and level of circumstellar interaction observed in other wavelength regions.

\subsection{Long-Term Mid-IR LCs}\label{sec:res_mir}

\begin{figure*}
\begin{center}
\includegraphics[width=15cm]{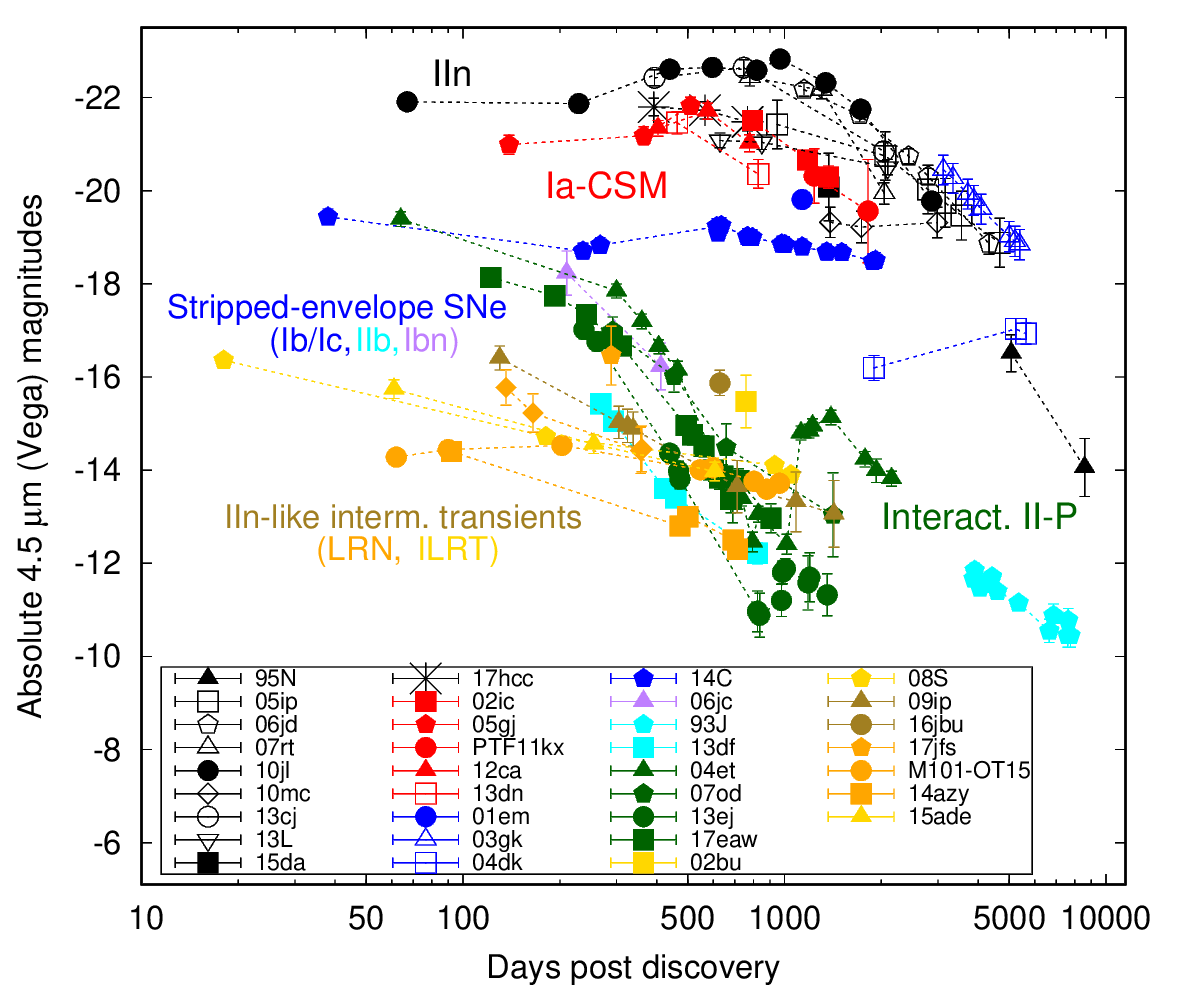}
\caption{{\it Spitzer} 4.5\,$\mu$m evolution of various types of interacting SNe. 
Sources of data are \citet{szalai19} and references therein, \citet{tinyanont19b}, \citet{jencson19}, and this work, as also highlighted in the text and in Table \ref{tab:phot}. Filled symbols denote SNe whose absolute magnitudes were determined with background subtraction (using either pre-explosion or very late-time reference images), while empty symbols, crosses, and asterisks denote objects where no background subtraction was possible.}
\label{fig:mir}
\end{center}
\end{figure*}

In Figure~\ref{fig:mir}, we present 4.5\,$\mu$m LCs of all SNe in our current sample and of some further well-known interacting SNe, along with their corresponding {\it Spitzer} archival data previously published by \citet[][and references therein]{szalai19}. 

\begin{figure*}
\begin{center}
\includegraphics[width=.8\linewidth]{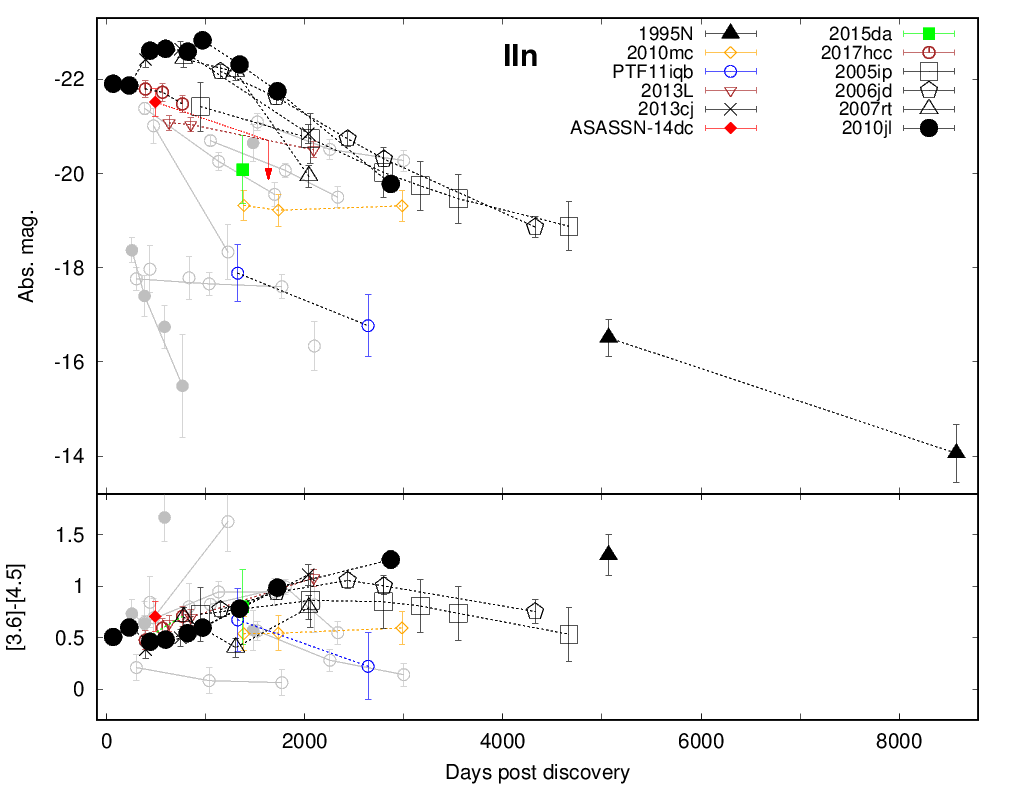}
\includegraphics[width=.8\linewidth]{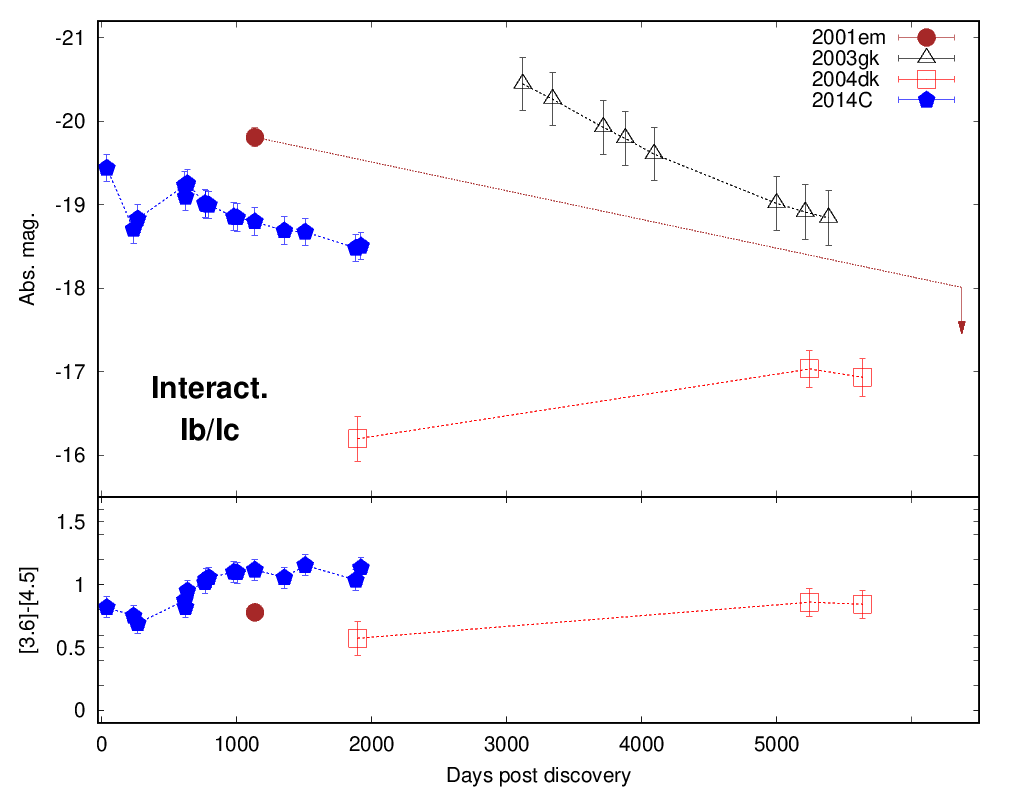}
\caption{4.5\,$\mu$m absolute magnitudes of all SNe~IIn and interacting SNe~Ib/c ever observed by {\it Spitzer} (data points are the same as in Fig. \ref{fig:mir}, but graphs are zoomed-in and time scales are linear), together with $[3.6]-[4.5]\,\mu$m color curves (all in Vega magnitudes). LASTCHANCE and some further well-sampled objects are highlighted, while all other published detections \citep[adopted from][]{szalai19} are marked with gray symbols. Existing nondetections (upper limits) are marked with arrows. Note that no 3.6\,$\mu$m measurements were obtained for the Type Ib SN 2003gk (bottom panel).}
\label{fig:mircolor}
\end{center}
\end{figure*}

The newest data originate primarily in this work, although some additional data from \citet{jencson19} and \citet{tinyanont19} are also included (see details in Secs.~\ref{sec:res_mir_ibc} and \ref{sec:res_mir_lowl}).  We also present the 4.5\,$\mu$m LCs of our SNe in Figure~\ref{fig:mircolor} for four subgroups: SNe~IIn, SNe~Ia-CSM, SNe~Ib/Ic, and intermediate-luminosity transients. 

\begin{figure*}
\begin{center}
\includegraphics[width=.48\linewidth]{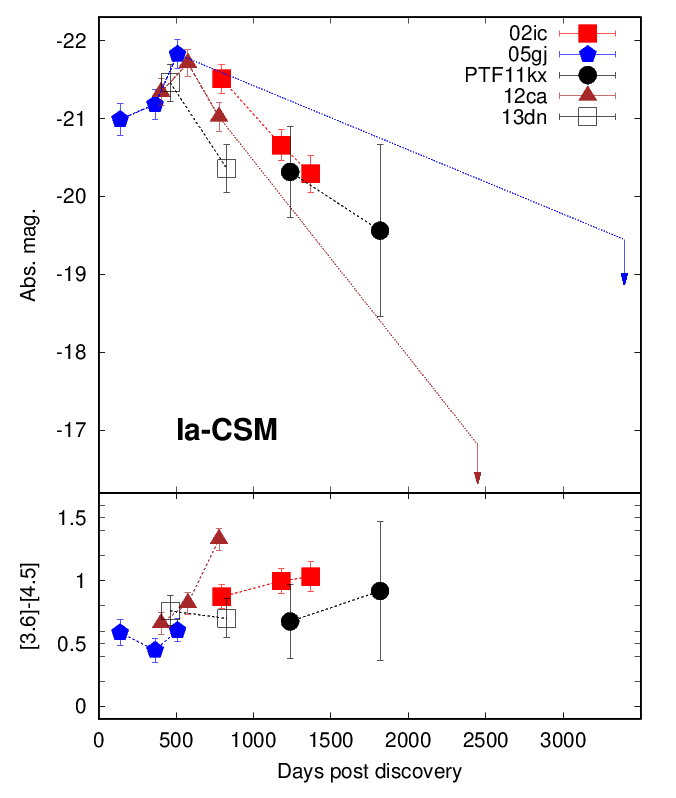}
\includegraphics[width=.48\linewidth]{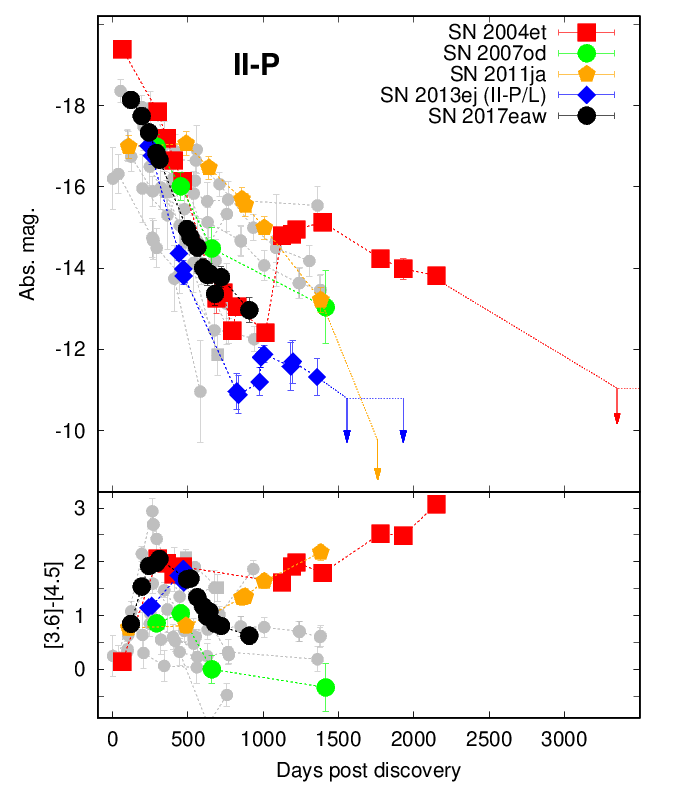}
\includegraphics[width=.48\linewidth]{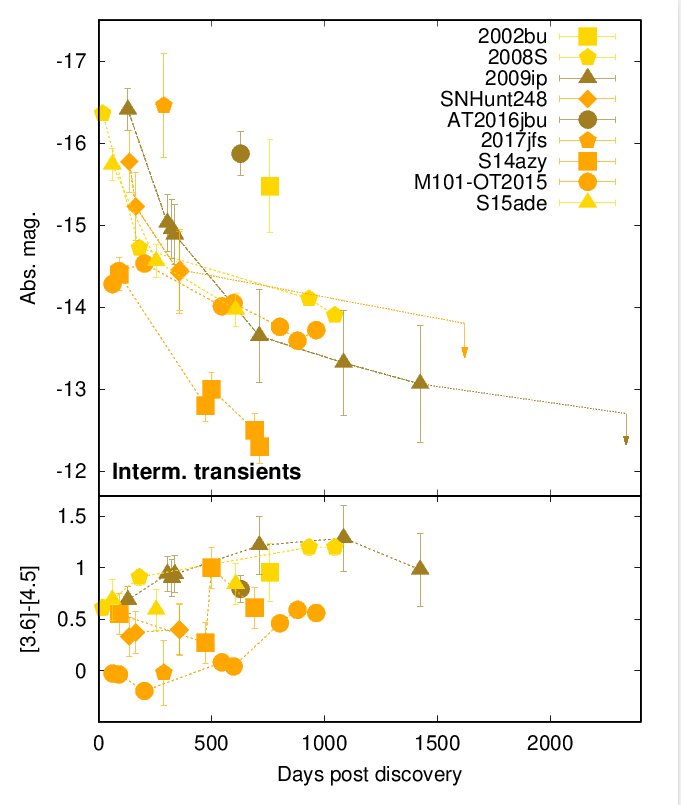}
\caption{4.5\,$\mu$m absolute magnitudes of all SNe~Ia-CSM ({\it top left}), SNe~II-P ({\it top right}), and intermediate-luminosity objects ({\it bottom}) ever observed by {\it Spitzer}. Note that there are missing positive 3.6\,$\mu$m detections for some of the objects. LASTCHANCE and some further well-sampled objects are highlighted, while all other published detections \citep[adopted from][]{szalai19} are marked with gray symbols. Existing nondetections (upper limits) are marked with arrows.}
\label{fig:mircolor2}
\end{center}
\end{figure*}


\subsubsection{Type IIn SNe}\label{sec:res_mir_iin}

In Figure~\ref{fig:mircolor}, existing mid-IR LCs of all SNe~IIn are presented; LASTCHANCE targets and some further well-sampled objects are highlighted. SNe~IIn form a bright but somewhat heterogeneous group in the mid-IR. Some extremely luminous objects reach a maximum brightness of $\sim -22$\,mag at 4.5\,$\mu$m, such as SN~2010jl \citep{andrews11a,fransson14,szalai19}, SN~2006jd \citep{fox10,stritzinger12,szalai19}, SN~2007rt \citep{fox11,fox13,szalai19}, and SN~2013cj \citep{szalai19}. Other SNe~IIn are much fainter in the mid-IR (i.e., SN PTF11iqb). Some objects even fall into the brightness gap that seems to exist between the previously mentioned objects (i.e., SN~2010mc).

Most SNe~IIn are still bright in our most recent {\it Spitzer} observations, in some cases $\sim 4500$--5000\,d (12--13\,yr!) after explosion.  The one SN~IIn exception to this is ASASSN-14dc, which is not detectable at $\sim 1600$\,d (but it is a very distant object, $d \approx 180$\,Mpc). We note that SN~1995N is still barely detectable in the 4.5\,$\mu$m image obtained during our LASTCHANCE survey at an age of $\sim 8600$\,d (making it the latest-observed SN~IIn in the mid-IR).

The decline rates of the mid-IR LCs of SNe~IIn vary from $\sim 0.4$\,mag\,(1000\,d)$^{-1}$ up to $\sim 3.0$\,mag\,(1000\,d)$^{-1}$; see Table \ref{tab:decrate}. If we compare these values with the mid-IR LC evolution of the objects (Figs.~\ref{fig:mir} and \ref{fig:mircolor}), we can see that the brightest ones (SNe~2007rt, 2010jl, and 2013cj) exhibit a faster decline after $\sim 1000$\,d. In the case of SN~2010jl, comprehensive multiwavelength studies and detailed analyses on ongoing dust formation exist \citep[e.g.,][]{andrews11a,gall14,fransson14,sarangi18,bevan20}; unfortunately, there are no similar datasets in the cases of SN~2007rt and SN~2013cj. Another group of SNe~IIn (SNe~2005ip, 2006jd, and 2013L) seems to be a bit fainter during the first $\sim 2000$\,d (note, however, that in the first two cases there are no mid-IR data before $\sim 950$\,d and $\sim 1150$\,d, respectively), but decline more slowly thereafter. In these SNe, the most probable scenario is a large, pre-existing dust shell that is continuously heated by energetic photons generated by ongoing CSM interaction \citep[see, e.g.,][]{fox10,fox11,fox20,stritzinger12,andrews17,taddia20}.
Based on this picture, the decline rates of these SNe may hint that the level of CSM interaction continuously decreases in the last years (except in the case of SN~2010mc, which seems to be in a very long ``plateau" phase even at $\sim 3000$\,d).


SN~2017hcc stands out among our LASTCHANCE sample because of its unusually high degree of polarization \citep{mauerhan17,kumar19} and the post-shock dust formation assumed from the analysis of its strongly blueshifted line profiles \citep{smith_ja20}. The SN was observed at three epochs (392, 565, and 767\,d) by {\it Spitzer}.
While at the earliest epoch, SN~2017hcc resembles SN~2010jl in the mid-IR, its fluxes do not increase after that but start to slowly decrease, similar to the LC of (for example) SN~2005ip. 

\begin{deluxetable}{lcccc}
\tablecaption{Decline rates of Type IIn SNe. \label{tab:decrate}}
\tabletypesize{\small}
\tablehead{
\colhead{Object} & \colhead{Range of epochs$^a$} & \colhead{$\Delta m_{3.6}$} & \colhead{$\Delta m_{4.5}$} \\
\colhead{} & \colhead{(d)} & \colhead{(1000\,d)$^{-1}$} & \colhead{(1000\,d)$^{-1}$}
}
\startdata
SN~1995N & 5067$-$8577 & --$^b$ & 0.7 \\
SN~2005ip & 2796$-$4667 & 0.4 & 0.6 \\
SN~2006jd & 2433$-$4525 & 0.8 & 1.0 \\
SN~2007rt & 1304$-$2042 & 3.5 & 3.0 \\
SN~2010jl & 969$-$2869 & 2.0 & 1.7 \\
SN~2013L & 850$-$2091 & 0.7 & 0.4 \\
SN~2013cj & 747$-$2036 & 1.8 & 1.4 \\
\enddata
\tablecomments{$^a$Used to calculate decline rates at the latest phases. $^b$SN~1995N was not detectable at 3.6\,$\mu$m on day 8577.}
\end{deluxetable}

\subsubsection{Type Ia-CSM SNe}\label{sec:res_mir_iacsm}

SNe~Ia-CSM are comparably luminous to SNe~IIn.  This subclass is thought to arise from thermonuclear explosions surrounded by dense, H-rich shells of ambient CSM \citep[producing SN~IIn-like emission features in their late-time spectra; see, e.g.,][]{silverman13,fox15,inserra16}. Few SNe~Ia-CSM have been observed by {\it Spitzer}; Figures~\ref{fig:mir} and \ref{fig:mircolor2} show all of them. Analysis of this small sample suggests that this subclass, particularly in the mid-IR, is more homogeneous than SNe~IIn \citep{graham17,szalai19}. Only SN~2012ca is observed around the mid-IR ``peak,'' reaching an absolute magnitude of almost $-22$ at 4.5\,$\mu$m around 600\,d after explosion, but the other objects in this subclass do seem to follow a qualitatively similar evolution. Except SN~2013dn, we also do not detect any of the SNe Ia-CSM between 1500\,d and 3500\,d after explosion.

We also discuss two nondetections of SN~2018fhw (ASASSN-18tb), the only new thermonuclear SN~Ia in our sample. This object has unique optical properties that show some similarities but also striking differences to other SNe~Ia-CSM \citep{kollmeier19,vallely19}. We observed the site of SN~2018fhw at two epochs ($\sim 250$\,d and $\sim 450$\,d after explosion). While there is a point-like source in all of the images, its position does not exactly match that of the SN \citep[its center is $\sim 2$\arcsec\ from the position given by][]{vallely19}; moreover, it does not show flux changes (within the uncertainties) between the two epochs at any channels. Given the lack of pre-explosion template images, we conclude for the moment that the source is not detected in our data. If we do calculate the absolute magnitudes of the detected mid-IR source, we obtain $\sim -17$\,mag in both IRAC channels, which is a much lower value than that of other studied SNe~Ia-CSM at these epochs. 

\begin{figure*}
\begin{center}
\includegraphics[width=.9\textwidth]{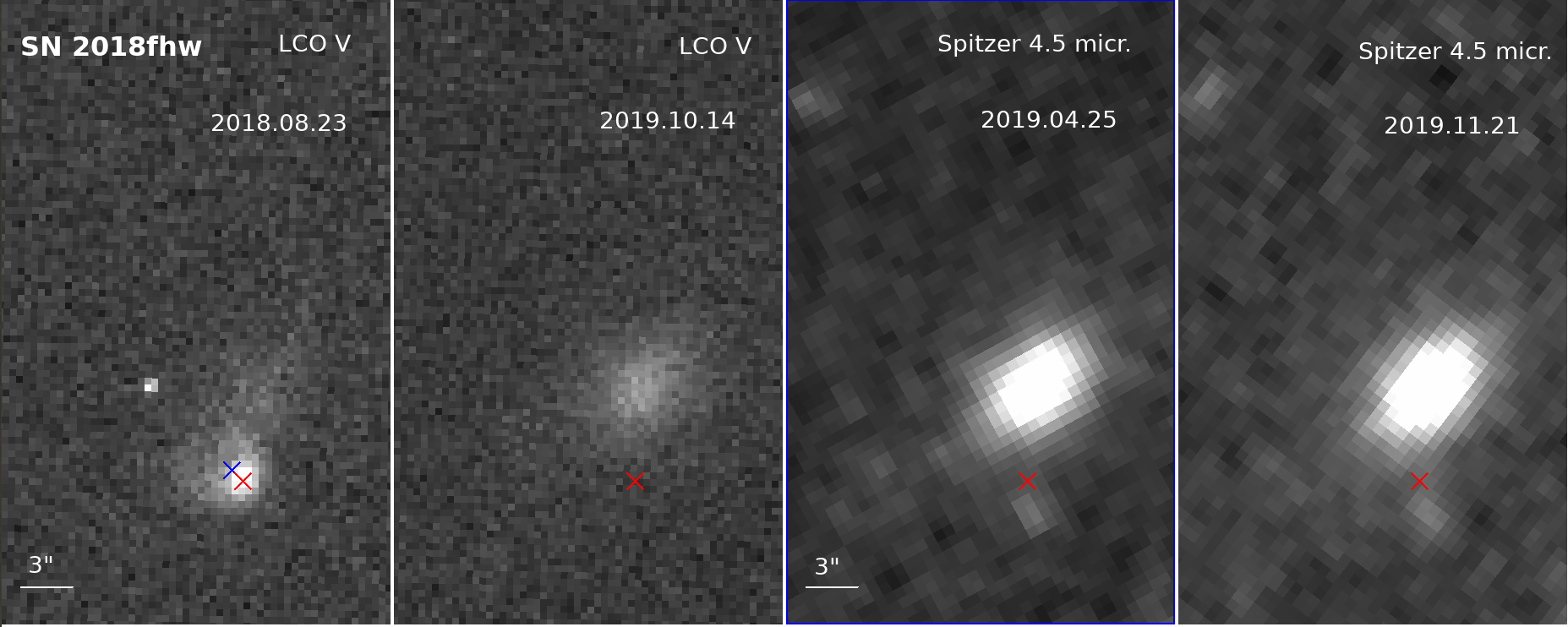}
\caption{Optical (LCO $V$-band) and mid-IR ({\it  Spitzer} 4.5\,$\mu$m) images of SN~2018fhw. The blue cross (in the left panel) denotes the coordinates given by \citet{vallely19}, while the red cross shows the photometric center of the optical (LCO) image of the source.} \label{fig:18fhw}
\end{center}
\end{figure*}



\subsubsection{Stripped-Envelope SNe}\label{sec:res_mir_ibc}

Given their lack of significant CSM, stripped-envelope (SE) SNe (i.e., SNe~Ibc) usually have uneventful mid-IR histories, producing a relatively quick fading and a disappearing after several hundred days.  Some SNe~IIb are detected in the mid-IR up to almost 1000\,d, which may be the sign of some (moderate) late-time CSM interaction; see \citet{szalai19} and references therein. Note that the Type IIb SN~1993J stands out owing to its proximity (3.5\,Mpc) and uncomplicated local background, detectable in IRAC imaging up to almost 8000\,d \citep{tinyanont16}.

There is now more than one known ``normal'' SE~SN that has been observed to transform into a strongly interacting, SN~IIn-like object at later times. The most well-studied event is the nearby ($d \approx 13$\,Mpc) Type Ib 
SN~2014C \citep{milis15,margutti17}. This object is also among the most well-sampled SNe with {\it Spitzer}, since its host galaxy is NGC 7331, one of the target galaxies for the SPIRITS program \citep{tinyanont16,tinyanont19}. SN~2014C was followed out to $\sim 2000$\,d (until the end of the {\it Spitzer} mission). During that time, it showed a clear rebrightening at $\sim 250$\,d as the CSM interaction began, and a very slow fading thereafter (it is still a bright source in the latest images).

Another unique SN~Ib/c that shows CSM interaction is SN~2001em. It had only one previous epoch of {\it Spitzer} data in 2004.  Increasing CSM interaction was detected at optical, X-ray, and radio wavelengths at that point \citep[see, e.g.,][]{pooley04,soderberg04,stockdale04}. While the object appears as a very bright source in the original imaging, our most recent nondetection provides a useful template (at 3.6\,$\mu$m and 4.5\,$\mu$m) that results in improved and updated mid-IR fluxes of SN~2001em at $\sim 1100$\,d \citep[which are $\sim 20$--25\% lower than those published by][]{szalai19}.  Figure~\ref{fig:mircolor} shows that SN~2001em was even more luminous in the mid-IR than SN~2014C, almost as bright as most SNe~IIn or SNe~Ia-CSM.

While it was not possible to carry out the same process on 5.8 and 8.0\,\micron\ data, SN~2001em appears as a very bright point source in those images, too. We concur that the previous conclusion of \citet{szalai19} still holds true, namely that (i) a two-component mid-IR spectral energy distribution (SED) in 2004 indicates the presence of multiple pre-explosion dust shells, and (ii) a few $10^{16}$\,cm for the shell radius and a total circumstellar dust mass of a few 0.01\,M$_{\odot}$, both of which (assuming a dust-to-gas mass ratio of 0.01) are in agreement with the results of \citet{chugai06} and \citet{chandra20}.

We also highlight the initially ``normal'' Type Ib/c SN~2004dk, which has started to show unexpected, strong signs of CSM interaction in the form of enhanced H$\alpha$ and X-ray emission more than a decade after its explosion \citep{mauerhan18,pooley19}. \citep[Note that in their earlier study,][already found some abrupt late-time radio variability indicating unusual circumstellar environment.]{wellons12} Although sampling is sparse, the mid-IR properties of SN~2004dk are consistent with these findings, showing a $\sim 1$\,mag increase between $\sim 2000$\,d and 5000\,d, subsequently slowly fading by the third epoch.  This is the latest-time mid-IR brightening observed in our sample.  Only SN~1987A has shown an increase in mid-IR emission at a later epoch, with continuously increasing mid-IR fluxes between $\sim 6000$\,d and 8500\,d \citep[][and references therein]{arendt20}.


Finally, we highlight SN~2003gk.  Although we do not have any recent observations, we include archival 4.5\,$\mu$m {\it Spitzer}/IRAC images uncovered during the writing of this paper. SN~2003gk is well sampled at 3.6\,\micron\ in the range 3100--5400\,d and shows a qualitatively similar evolution to that of the long-term monitored SNe~IIn 2005ip and 2006jd. There is only one publication on SN~2003gk in the literature \citep{bietenholz14}, which suggests the presence of late-time CSM interaction based on radio observations.

\subsubsection{Type II-P SNe}\label{sec:res_mir_II-P}

Based on theoretical expectations \citep[see, e.g.,][]{kozasa09,gall11}, SNe~II-P are likely the best candidates for (ejecta) dust formation among SNe.  Some of these objects were targets of {\it Spitzer} observations in the early years of the mission. These data typically trace dust formation $\sim 1$--3\,yr after explosion and estimate the physical parameters of newly-formed dust \citep[e.g.,][]{meikle07,meikle11,sugerman06,kotak09,andrews10,fabbri11,szalai11,szalai19,szalai13}. The results do not support either the theoretical prediction of significant($\gg0.001\,M_{\odot}$) SN dust production or the large observed dust masses in some young Galactic SN remnants and in high-redshift galaxies. This discrepancy can be partly resolved by the application of clumped dust models, or significant grain growth in the interstellar matter (ISM) \citep[see][for a detailed review]{gall11}. Another possibility is that a significant amount of cold ($<50$\,K) dust may be present in the ejecta, as can be seen via far-IR and sub-mm observations of the very nearby SN~1987A \citep{matsuura11,matsuura15,indebetouw14,wesson15}.

During our LASTCHANCE program, we collected single data points on some young SNe~II-P during the assumed formation period of ejecta dust ($\sim 200$--650\,d): SNe 2017aym, 2017eaw, 2017ivu, and 2018acj. Moreover, we also targeted the older SNe~II-P 2012aw and 2013ej, both of which turned to be below the detection limit. 
Regarding SN~2013ej \citep{mauerhan17}, a very late-time mid-IR rebrightening can be seen between $\sim 700$ and 1000\,d, just as in the SN~II-P 2004et \citep{kotak09,fabbri11}, presumably because of dust formation in the CDS behind the reverse shock and not within the ejecta. Based on known \citep{tinyanont19b,szalai19b} and previously unpublished {\it Spitzer} data on SN~2017eaw, it seems to produce a ghost-like similarity to SN~2004et (found in the same host galaxy!) in the mid-IR (even the small rebrightening is there at $\sim 700$--750\,d; see Fig.~\ref{fig:mircolor2}). Unfortunately, {\it Spitzer}'s mission ended just before the expected intense rebrightening seen in the case of SN~2004et and SN~2013ej at $\sim 1000$--1300\,d (the last {\it Spitzer} observation of SN~2017eaw was obtained in Nov. 2019). Nevertheless, connecting to the similarity of these three objects and the main storyline of our current paper, all of them show signs of late-time circumstellar interaction at $\sim 900$\,d observed via the emerging H$\alpha$ emission-line profiles \citep[see][]{weil20}.

\subsubsection{Intermediate-Luminosity Interacting Transients}\label{sec:res_mir_lowl}

In Figures~\ref{fig:mir} and \ref{fig:mircolor2}, we plot both previously published and new mid-IR data points of intermediate-luminosity interacting transients. In the literature, this subclass is divided into two groups: intermediate-luminosity red transients (ILRTs) and luminous red novae (LRNe). ILRTs are typically described as explosions of deeply dust-enshrouded stars \citep[see, e.g.,][]{prieto08,bond09,kochanek11,jencson19}.  Included in our sample are SNe~2008S \citep{szczygiel12a}, 2002bu \citep{szczygiel12b}, and SPIRITS15ade \citep{jencson19}. LRNe are thought to be merging events of massive binaries \citep[see, e.g.,][and references therein]{pastorello19b}.  We included data for SNHunt248 \citep{mauerhan18b}, M101-2015OT-1 \citep{blagorodnova17,jencson19}, SPIRITS14azy \citep{jencson19}, as well as the previously unpublished data of SN~2017jfs.


These objects are less luminous in the mid-IR than normal SNe 
and show a significant heterogeneity in their luminosity:
at $\sim 300$\,d and $\sim 600$--700\,d, when there are data points for most of these SNe, brightness values spread over nearly a $\sim 3$\,mag range. 
Most of well-sampled transients of that kind fade quickly, but M101-OT2015-1 shows a plateau at $\sim 800$--960\,d. These differences likely arise from different physical and/or geometric properties of the ambient dust content.

We also include {\it Spitzer} data on the well-studied SN~2009ip. Our most recent observation is a nondetection, allowing for template-based background subtraction. Our new photometry is within the original uncertainties \citep[][]{fraser15,szalai19}. 
It is worth noting that while SN~2009ip reached an optical ($R$-band) peak brightness similar to that of normal SNe~IIn \citep[$\sim -$18\,mag; see, e.g.,][]{margutti14} during its outburst in 2012, it is an order of magnitudes fainter in the mid-IR than either Type IIn or other ``normal" interacting SNe. Based on the detailed analyses of \citet{margutti14}, \citet{fraser15}, and others, the quick optical fading of SN~2009ip implies that it was a low-energy explosion, while results of spectropolarimetric studies can be explained best with the presence of an inclined disk-like CSM \citep{reilly17}. These two effects together can explain the low, continuously decreasing mid-IR luminosity.

Finally, we highlight the single data point for AT2016jbu, which showed a spectral evolution similar to that of SN~2009ip \citep{fraser17,brennan21a,brennan21b}.
Its mid-IR brightness, however, is more similar to that of the ILRT-type SN~2002bu than of SN~2009ip at the epoch of our {\it Spitzer} observation. This case also suggests that the mid-IR properties of intermediate-luminosity interacting transients and, therefore, their local environment cannot be distinguished clearly by spectral type alone.

\subsection{Mid-IR Color Curves and Dust Temperatures}\label{sec:res_mircolor}

Figures~\ref{fig:mircolor} and \ref{fig:mircolor2} also show $[3.6]-[4.5]$\,\micron\ color curves in every case when data in both channels are available. Mid-IR color evolution can be, with some limitations, a good indication of temporal changes of dust temperature.  In the first few hundreds of days after explosion, when hot ($T \gtrsim 1000$\,K) components of gas and dust tend to dominate the ejecta and/or its environment, the IRAC measurements probe the peak of the dust blackbody emission and provide the best constraints on the physical parameters.  Such measurements have been illustrated with several SNe~II-P \citep[see, e.g.,][]{sugerman06,meikle07,kotak09,andrews10,andrews11b,szalai11,szalai13} and several interacting SNe (SN~2005ip -- \citet{fox10,stritzinger12,fox20}; SN~2006jd -- \citet{stritzinger12}; SN~2010jl -- \citet{andrews11a,fransson14,sarangi18,bevan20}; SN~2013L -- \citet{andrews17,taddia20}; SN~2014C -- \citet{tinyanont19}; SN~2015da -- \citet{tartaglia20}). 

Since most of the data presented in this paper were obtained during the {\it Warm Spitzer} mission, we are limited to the 3.6 and 4.5\,\micron\ data.  Longer-wavelength data would be necessary to probe (much) colder dust, which has been observed in SN~1987A \citep{dwek10,matsuura11,indebetouw14} and in young Galactic SN remnants \citep[see, e.g.,][for a review]{williams_temim17}.
A limited amount of 5.8--24.0\,$\mu$m measurements were obtained by {\it Spitzer} during its cryogenic phase on a few extragalactic SNe; from these data, presence of cold dust can be also inferred \citep[even within a few years after explosion, see][for a review]{szalai19}.
Unfortunately, the assumed (significant) cold dust reservoirs are not able to be efficiently surveyed; however, above $\sim$200 K, we will able to get a much detailed picture about the dust content of SNe 
due to the expected data of the forthcoming {\it James Webb Space Telescope (JWST)}.


Even with just two data points, we can derive dust temperatures and radii of the dust-containing regions via fitting blackbodies (or simple dust models) to two-point SEDs \citep[see, e.g.,][]{fox11,fox13,szalai19}.  First, it is important to determine if any line emission contributes to the 3.6 and 4.5\,$\mu$m measurements. As shown in some SNe~II-P  \citep[e.g.,][]{kotak05,szalai11,szalai13}, additional flux
from the 1--0 vibrational band of CO at 4.65\,$\mu$m may arise in the first $\sim 500$\,d, strongly affecting the 4.5\,$\mu$m LCs and leading to $[3.6]-[4.5] \approx 1.5$--3\,mag values in this period (see e.g. SNe 2004et and 2017eaw in Fig.~\ref{fig:mircolor2}).

The colors of SNe~IIn have a relatively small scatter ($\sim 0.5$--0.7\,mag), indicating the long-time presence of hot dust above $\sim 1000$\,K. This is in agreement with the results of \citet{fransson14} and \citet{sarangi18} on SN~2010jl and of \citet{kokubo19} on KISS15s. The color of SN~2010jl turns continuously redder after $\sim 1000$\,d, probably indicating the decreasing dust temperature, which is also observed in SN~2007rt and SN~2013cj (at least up to their last observed epochs). In contrast, the colors of SN~2005ip and SN~2006jd indicate nearly constant (or, at least, slowly changing) dust temperatures out to $\sim 4500$\,d, in agreement with the results of \citet{fox10}, \citet{stritzinger12}, \citet{fox13}, and \citet{fox20}.
SN~2010mc seems to follow a similar trend as SN 2005ip. 


SNe~Ia-CSM and interacting SNe~Ib/c, basically, show a nearly constant or slowly changing mid-IR color evolution within their observational ranges. This is in agreement with the results of previous studies \citep[e.g.][]{FF13,graham17,tinyanont19}, in which authors revealed slow evolution of dust temperatures in some of these SNe.
The only exception is SN~2012ca shows a more intense reddening (cooling) in the first few hundred days. 

The colors of interacting intermediate-luminosity transients, just as their LCs, exhibit large heterogeneity. The mid-IR color evolution of most of these objects has been described in detail by \citet{jencson19}. Regarding our new targets, AT2016jbu and SN~2017jfs have been observed at only one epoch, while SNHunt248 -- captured in total four times -- shows a nearly constant, relatively blue color between 136 and 360\,d.

\subsection{A Comprehensive Multiwavelength Overview of LCs of Interacting SNe}\label{sec:res_mw}

\begin{table*}
\begin{center}
\caption{\label{tab:multidat} Presence of late-time multiwavelength datasets for LASTCHANCE targets and other interacting SNe}
\newcommand\T{\rule{0pt}{3.1ex}}
\newcommand\B{\rule[-1.7ex]{0pt}{0pt}}
\begin{tabular}{l|ccccc|c}
\multicolumn{1}{c|}{Object} & Mid-IR & Near-IR & H$\alpha$-flux & X-ray & Radio & Refs.\T\B \\
\hline
\hline
\multicolumn{1}{c|}{\bf IIn} & \multicolumn{6}{c}{}\T\B \\
\hline
SN~1995N & Yes & Yes & -- & Yes & Yes & 1-9\T \\
SN~2005ip & Yes & Yes & Yes & Yes & Yes & 1, 10-16 \\
SN~2006jd & Yes & Yes & Yes & Yes & Yes & 1, 13, 16-17 \\
SN~2010jl & Yes & Yes & Yes & Yes & Yes & 1, 16, 18-21 \\
PTF11iqb & Yes & -- & Yes & -- & -- & 1, 22-23 \\
SN~2013L & Yes & Yes & Yes & -- & -- & 1, 24, 25 \\
SN~2015da & Yes & Yes & Yes & -- & -- & 1, 26 \\
KISS15s & Yes & Yes & Yes & -- & -- & 27\B \\
\hline
\multicolumn{1}{c|}{\bf IIn/impost.} & \multicolumn{6}{c}{}\T\B \\
\hline
SN~2009ip & Yes & Yes & Yes & -- & -- & 1, 16, 23, 28-30\T\B \\
\hline
\multicolumn{1}{c|}{\bf Ib/c} & \multicolumn{6}{c}{}\T\B \\
\hline
SN~2001em & Yes & -- & -- & Yes & Yes & 1, 16, 31-36\T \\
SN~2003gk & Yes & -- & -- & -- & Yes & 1, 37 \\
SN~2004dk & Yes & -- & Yes & -- & Yes & 1, 23, 38-40 \\
SN~2014C & Yes & Yes & Yes & Yes & Yes & 39, 41-44\B \\
\hline
\multicolumn{1}{c|}{\bf Ia-CSM} & \multicolumn{6}{c}{}\T\B \\
\hline
SN~2012ca & Yes & Yes & -- & Yes & -- & 1, 16, 45-47\T \\
SN~2013dn & Yes & Yes & -- & -- & -- & 16, 45\B\\
\hline
\multicolumn{1}{c|}{\bf II-P} & \multicolumn{6}{c}{}\T\B \\
\hline
SN~2004et & Yes & Yes & Yes & -- & -- & 48-51 \T \\
SN~2013ej & Yes & Yes & Yes & -- & -- & 16, 51-53 \\
SN~2017eaw & Yes & Yes & Yes & -- & -- & 1, 51, 54-55\B \\
\end{tabular}
\tablecomments{{\it References}:
$^1$This work;
$^2$\citet{fox00}; $^3$\citet{fransson02}; $^4$\citet{gerardy02}; $^5$\citet{zampieri05}; $^6$\citet{chandra05}; $^7$\citet{chandra09}; $^8$\citet{pastorello11}; $^9$\citet{VD13}; $^{10}$\citet{fox09}; $^{11}$\citet{fox10}; $^{12}$\citet{fox11}; $^{13}$\citet{stritzinger12}; 
$^{14}$\citet{katsuda14}; $^{15}$\citet{smith17}; $^{16}$\citet{szalai19}; $^{17}$\citet{chandra12}; 
$^{18}$\citet{andrews11a}; $^{19}$\citet{fransson14}; $^{20}$\citet{chandra15}; $^{21}$\citet{katsuda16}; 
$^{22}$\citet{smith15}; $^{23}$\citet{mauerhan18}; $^{24}$\citet{andrews17};
$^{25}$\citet{taddia20};
$^{26}$\citet{tartaglia20}; 
$^{27}$\citet{kokubo19};
$^{28}$\citet{graham14}; $^{29}$\citet{fraser15}; $^{30}$\citet{graham17}; $^{31}$\citet{pooley07}; $^{32}$\citet{stockdale07}; $^{33}$\citet{kelley07}; $^{34}$\citet{bietenholz07}; $^{35}$\citet{schinzel09}; $^{36}$\citet{chandra20};
$^{37}$\citet{bietenholz14}; $^{38}$\citet{wellons12}; $^{39}$\citet{vinko17}; $^{40}$\citet{pooley19}; $^{41}$\citet{tinyanont16}; $^{42}$\citet{margutti17}; $^{43}$\citet{tinyanont19}; $^{44}$\citet{bietenholz18};
$^{45}$\citet{fox15}; $^{46}$\citet{inserra16}; $^{47}$\citet{bochenek18};
$^{48}$\citet{kotak09}; $^{49}$\citet{fabbri11}; $^{50}$\citet{maguire10};
$^{51}$\citet{weil20}; $^{52}$\citet{mauerhan17}; $^{53}$\citet{dhungana16}.
}
\end{center}
\end{table*}

\begin{figure*}[ht]
\begin{center}
\includegraphics[width=.3\linewidth]{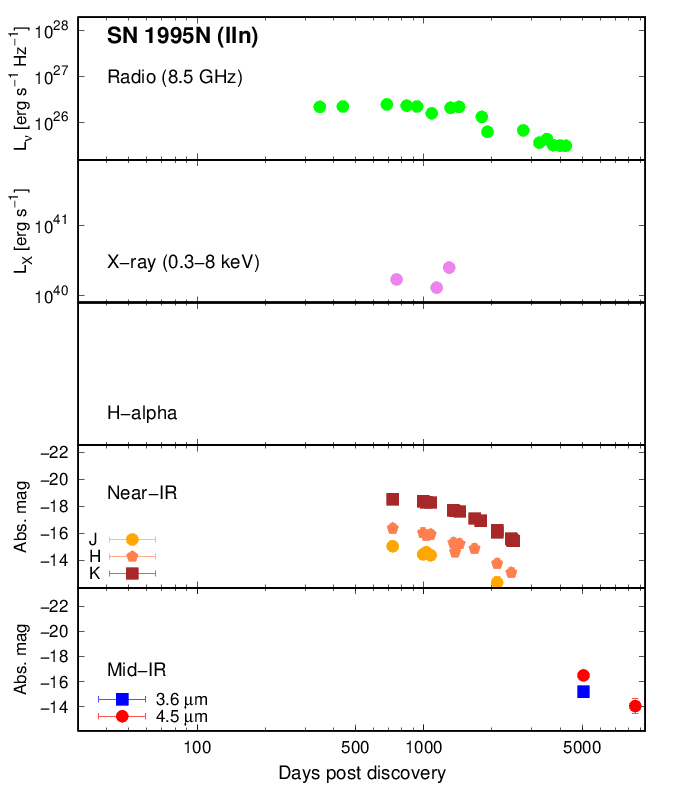}
\includegraphics[width=.3\linewidth]{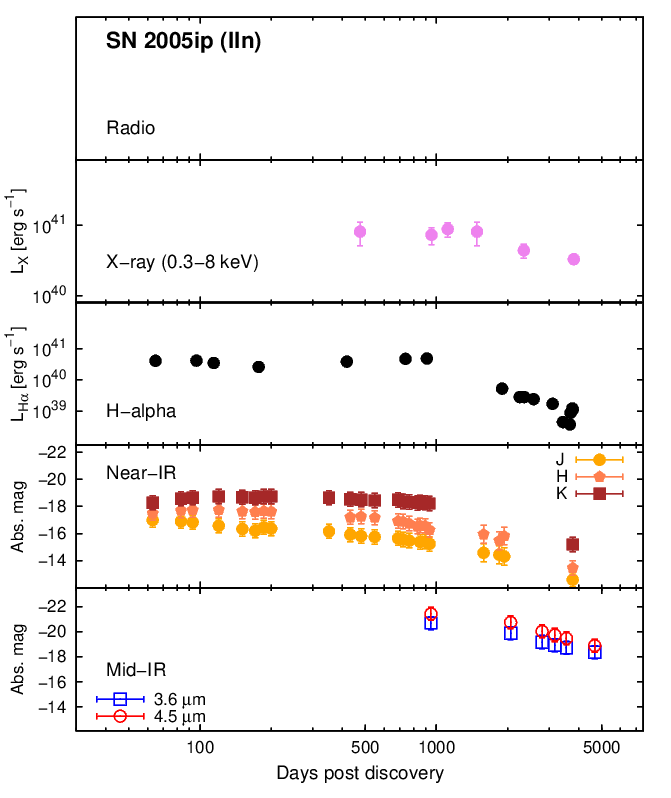}
\includegraphics[width=.3\linewidth]{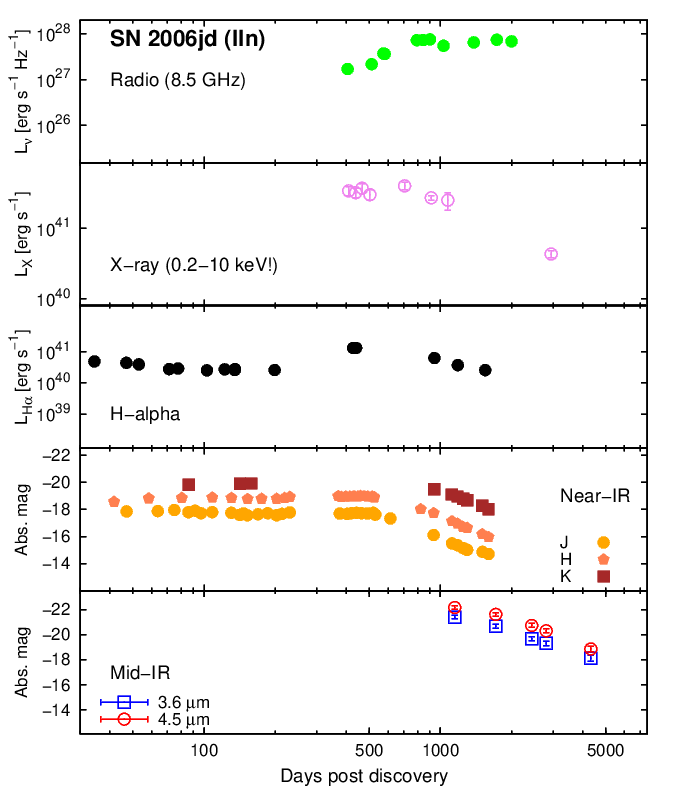}
\includegraphics[width=.3\linewidth]{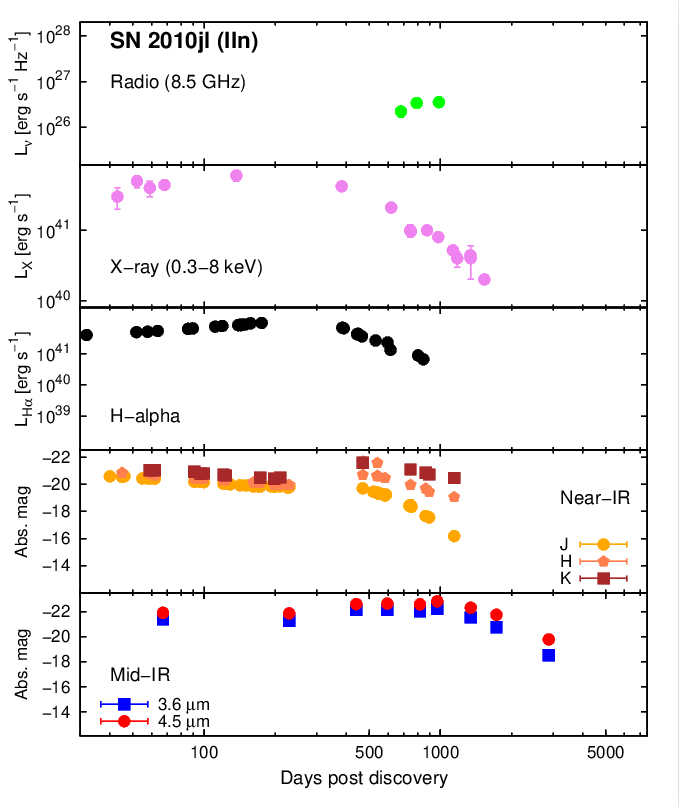}
\includegraphics[width=.3\linewidth]{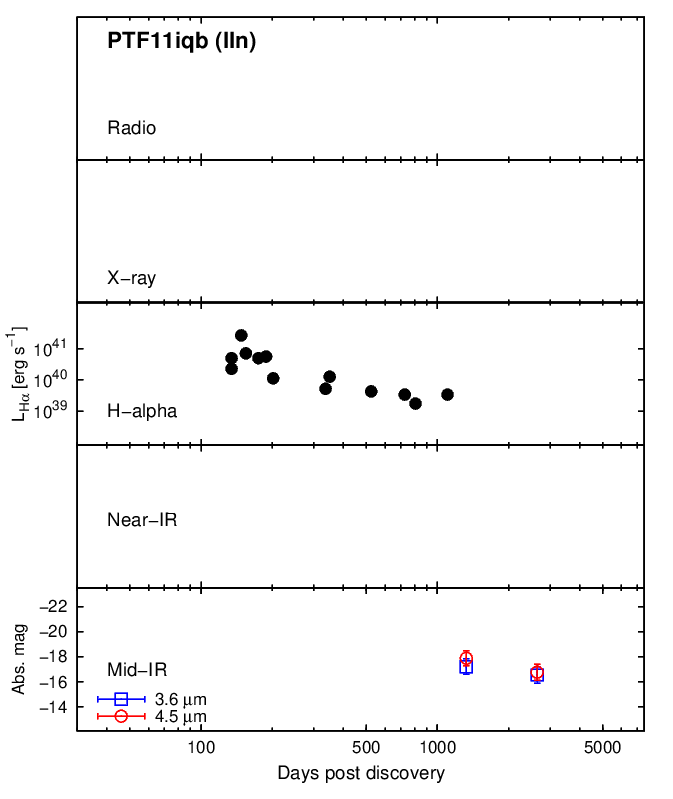}
\includegraphics[width=.3\linewidth]{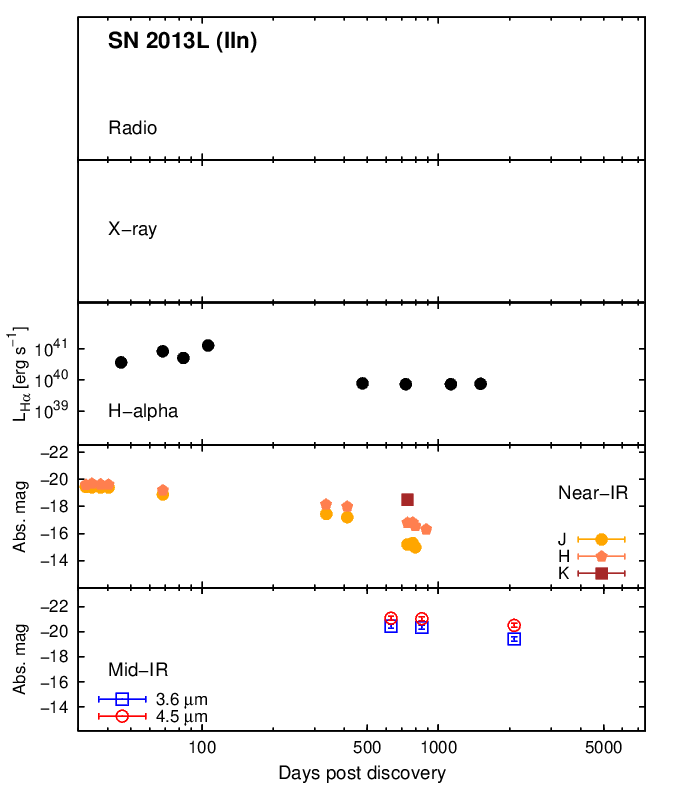}
\includegraphics[width=.3\linewidth]{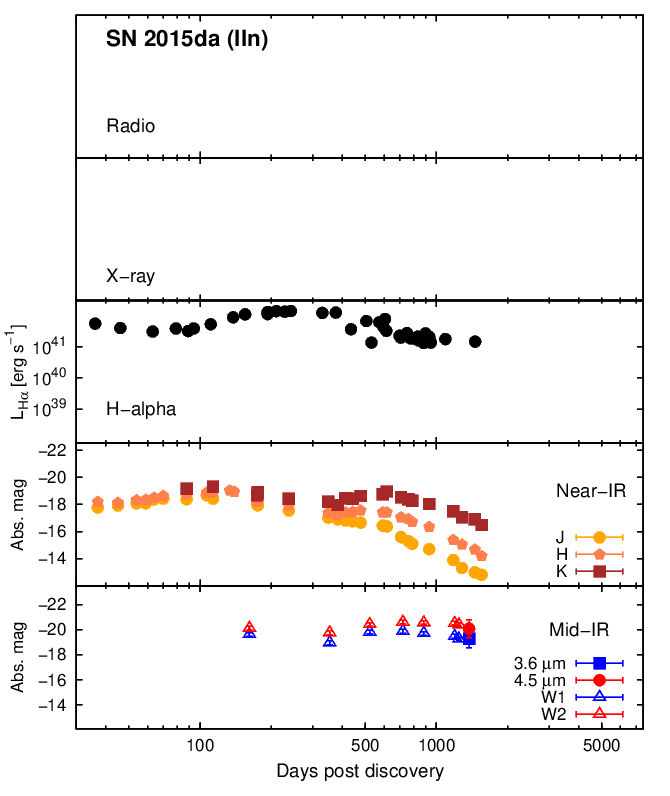}
\includegraphics[width=.3\linewidth]{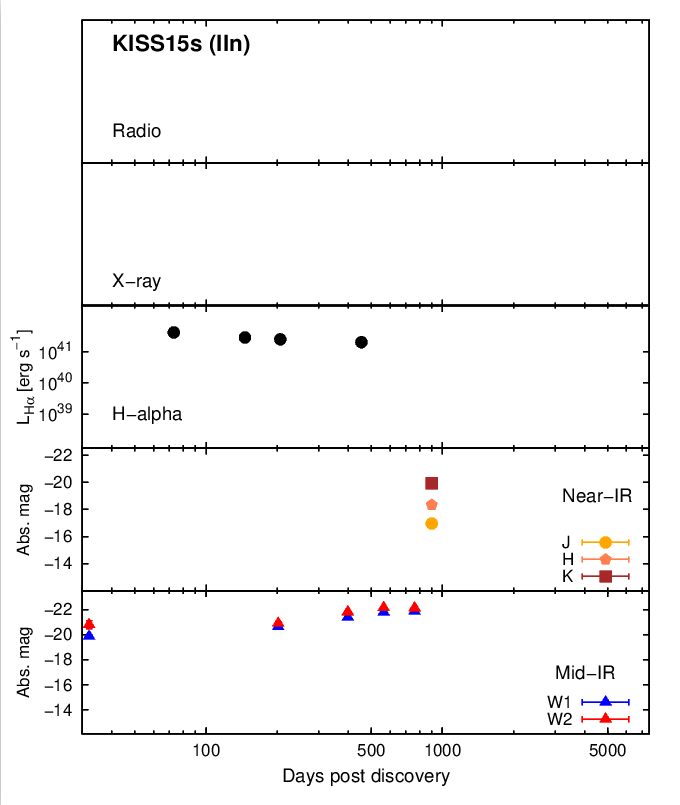}
\caption{Comparative figures showing multiwavelength LCs of SNe~IIn; sources of data are shown in Table \ref{tab:multidat}. Regarding mid-IR data, circles/squares and triangles denote {\it Spitzer} and {\it NEOWISE} data, respectively; filled and empty symbols denote values determined with or without background subtraction, respectively. Regarding X-ray data, filled circles denote unabsorbed luminosities measured in the 0.3--8\,keV regime, while empty circles denote values covering larger energy ranges.}
\label{fig:lc_IIn}
\end{center}
\end{figure*}

\begin{figure*}
\begin{center}
\includegraphics[width=.3\linewidth]{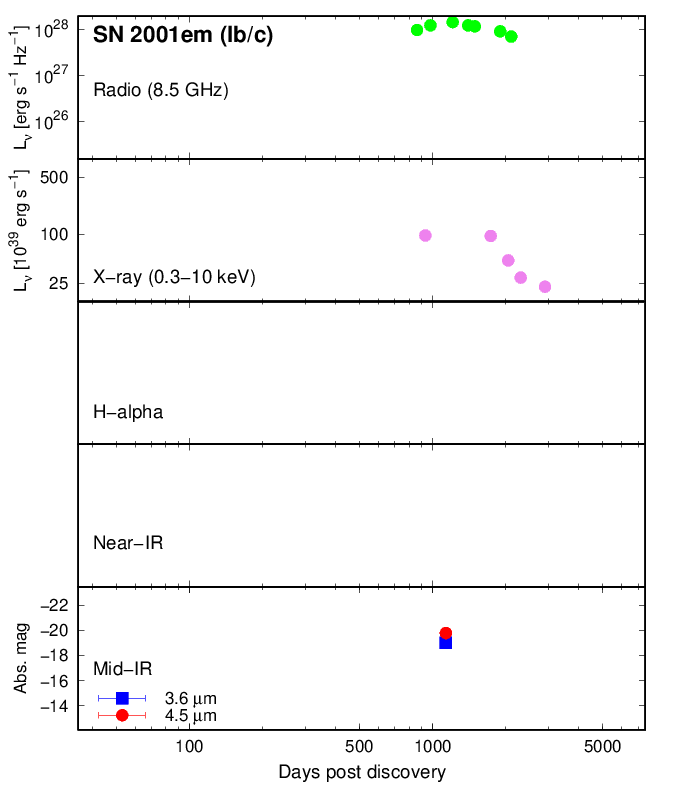}
\includegraphics[width=.3\linewidth]{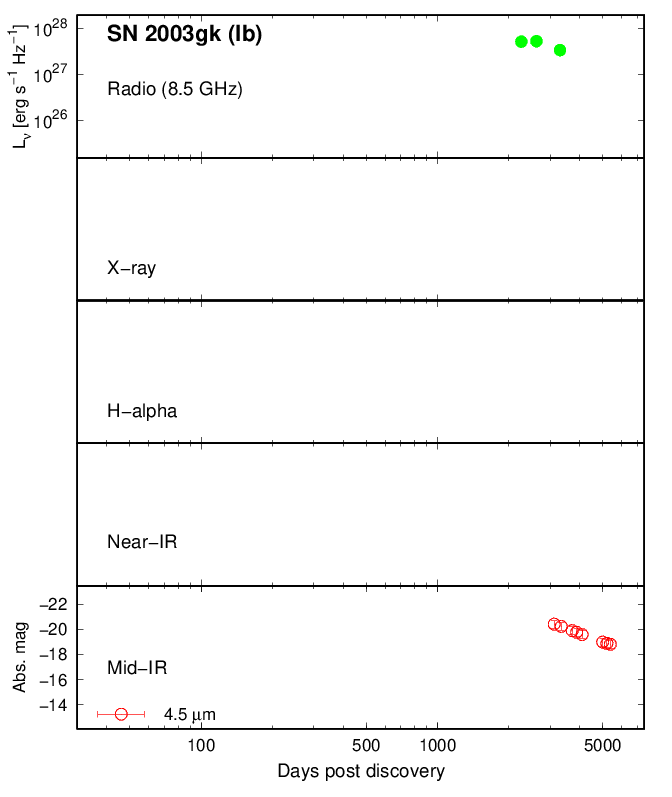}
\includegraphics[width=.3\linewidth]{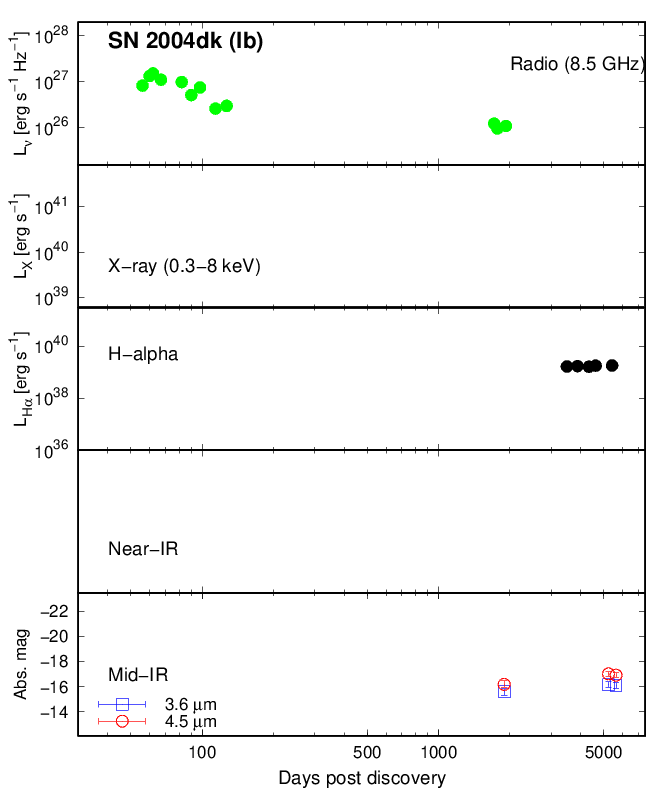}
\includegraphics[width=.3\linewidth]{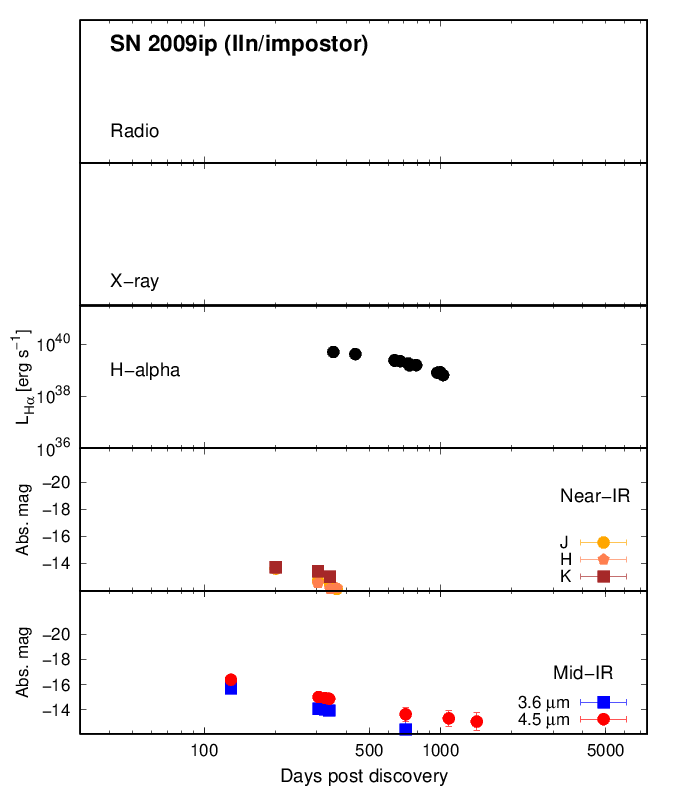}
\includegraphics[width=.3\linewidth]{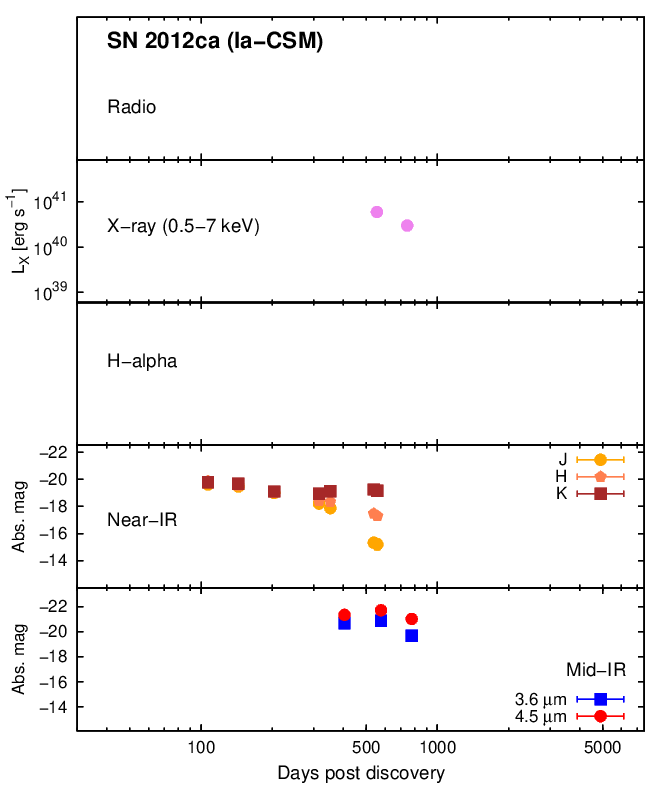}
\includegraphics[width=.3\linewidth]{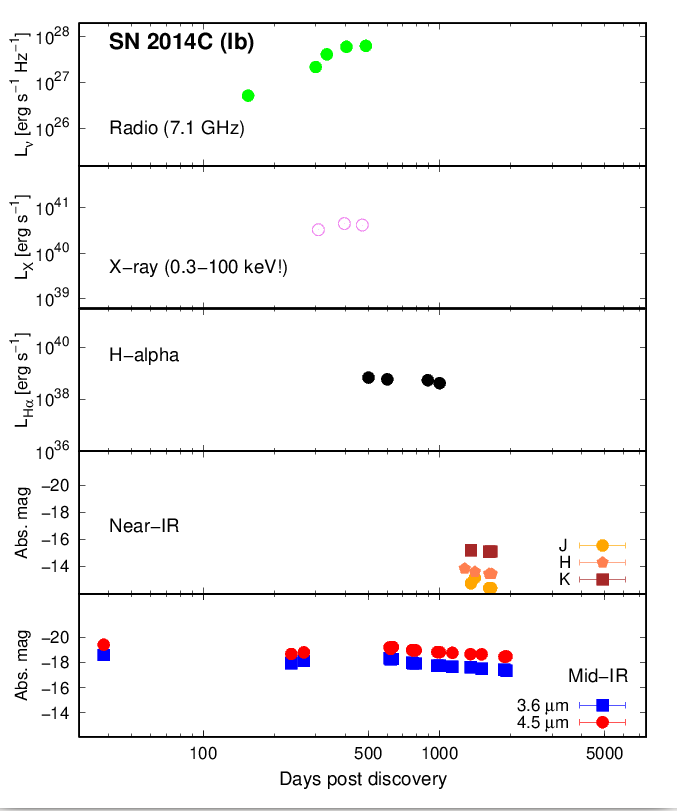}
\includegraphics[width=.3\linewidth]{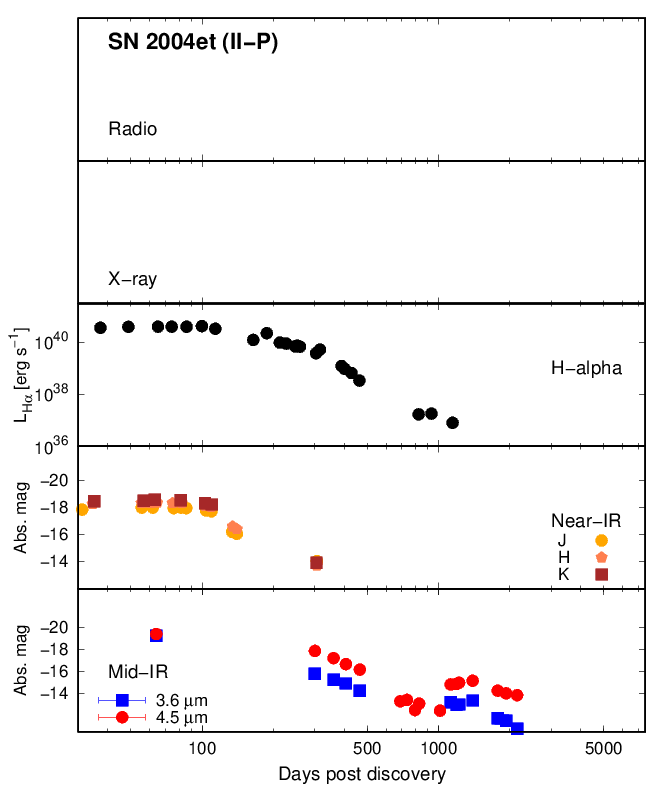}
\includegraphics[width=.3\linewidth]{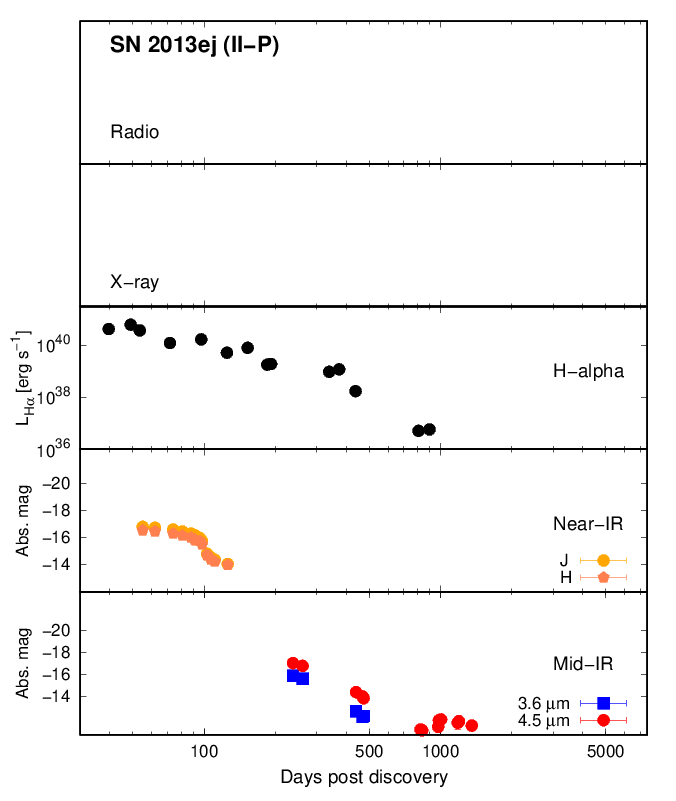}
\includegraphics[width=.3\linewidth]{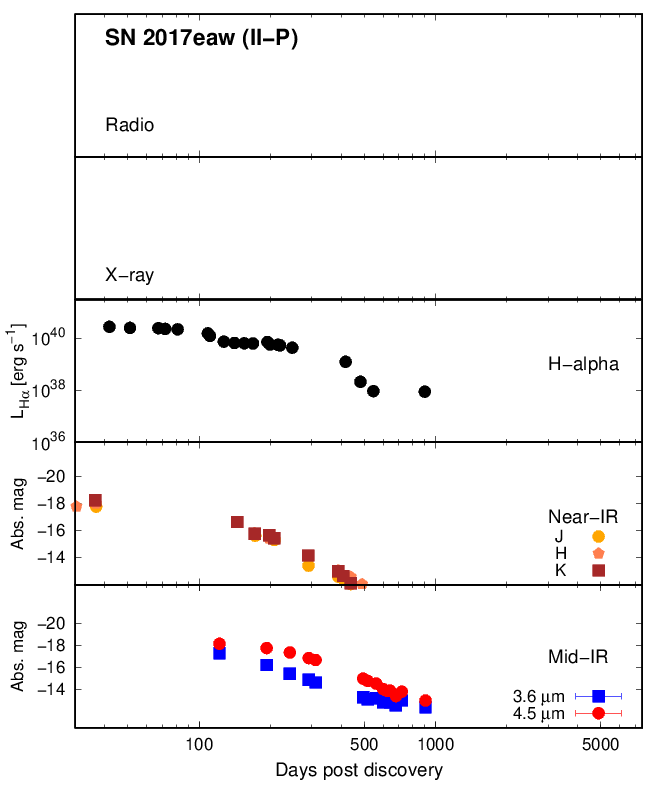}
\caption{Comparative figures showing multiwavelength LCs of further late-time interacting SNe; sources of data are shown in Table \ref{tab:multidat}. Regarding mid-IR data, filled and empty symbols denote values determined with or without image subtraction, respectively. Regarding X-ray data, filled circles denote unabsorbed luminosities measured in the 0.3--8\,keV regime, while empty circles denote values covering larger energy ranges.}
\label{fig:lc_I}
\end{center}
\end{figure*}

\begin{figure*}
\begin{center}
\includegraphics[width=.45\textwidth]{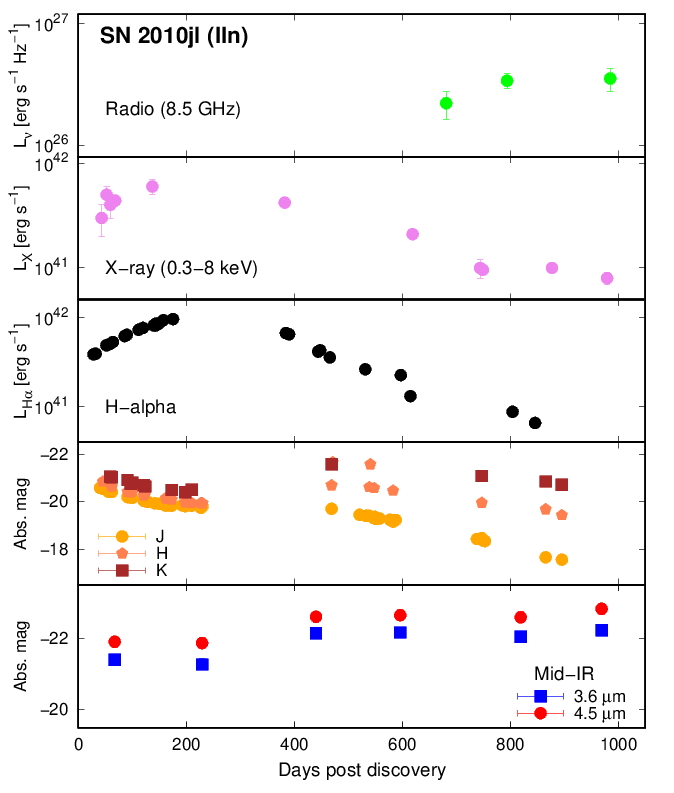}
\includegraphics[width=.45\textwidth]{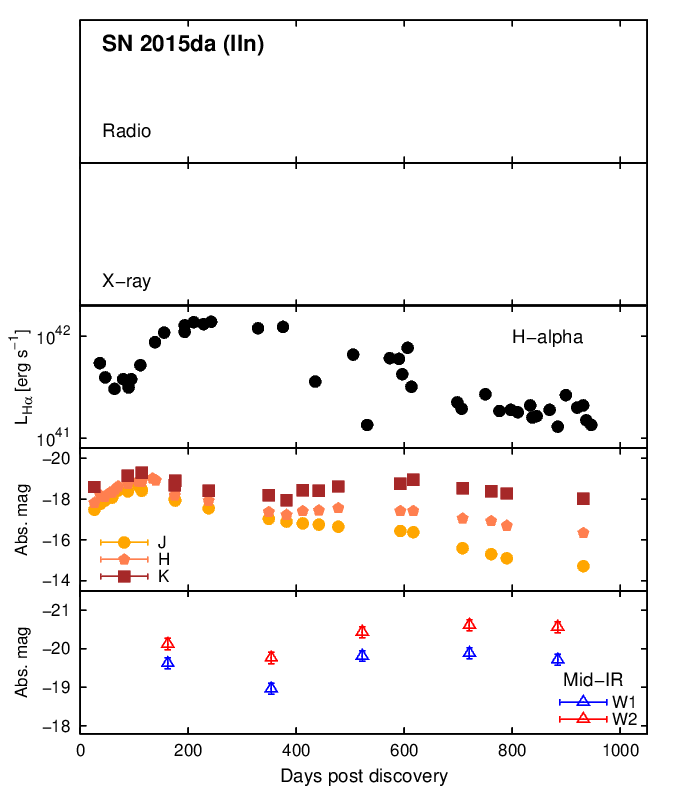}
\caption{The same as in Fig.~\ref{fig:lc_IIn}, but using a linear time scale and zoomed in the first $\sim 1000$\,d of SN~2010jl ({\it left}) and SN~2015da ({\it right}).}
\label{fig:lc_10jl}
\end{center}
\end{figure*}

Beyond the mid-IR, signs of ongoing circumstellar interaction can be most efficiently detected and traced in the form of radio, X-ray, or H$\alpha$ emission. \citet{fox11} proposed a scenario whereby pre-shocked dust shells are radiatively heated by ongoing shock interaction at the inner radii, but the exact relationship between the different wavelengths may not be straightforward.  Long-term multiwavelength monitoring, however, is quite limited; see recent reviews by \citet{cf17} and \citet{chandra18}, as well as (for example) \citet{weiler02}, \citet{dwarkadas12}, and \citet{vinko17} regarding radio, X-ray, and H$\alpha$ data, respectively.  


Figures~\ref{fig:lc_IIn} and \ref{fig:lc_I}, 
and Table \ref{tab:multidat}, show an overview of long-term multiwavelength (radio, X-ray, H$\alpha$, near-IR, and mid-IR) LC evolution of many SNe in our sample (as well as Fig.~\ref{fig:lc_10jl}, zooming into the early-time evolution of SNe 2010jl and 2015da; see discussion below).  While there are still only a few objects with well-sampled multichannel datasets, some conclusions can be drawn.

First, from the limited data available, near-IR and mid-IR LCs appear to have qualitatively similar evolution, particularly in the $K$ band, which suggests that the dominant source of the detected flux at these wavelengths is the same -- thermal continuum emission and not line emission. The availability of instruments with access to wavelengths longer than $\sim2$\,$\mu$m is expected to be limited for high-cadence follow-up observations of transient objects in the near-term post-{\it Spitzer} era. Long-term NIR ($K$-band) observations can be useful for monitoring the late-time evolution of the hottest dust when long-wavelength observations are unavailable. This could be particularly useful during the upcoming era of the {\it Nancy Grace Roman Space Telescope (NGRST)}, which will be usable for similar follow-up observations of SNe (its longest-wavelength filter, F184, covers 1.68--2.00 $\mu$m, falling between the traditional $H$ and $K$ bands). We note, however, that the fundamental vibrational overtones of CO may contribute to not only 4.5\,$\mu$m but also to $K$-band fluxes of certain SNe \citep[see, e.g.,][]{jencson17,jencson19}.

Furthermore, the average brightness levels of the objects seem to correlate with each other in every wavelength range (except radio). \citet{dwarkadas12} and \citet{snax17} published a compilation of X-ray LCs of different types of SNe, and their comparative figures show similarities to our mid-IR LC compilation (Fig.~\ref{fig:mir}) from the viewpoint that (some) SNe~IIn represent the brightest sources, while observed luminosities of other SNe scatter within a wide range. In the near-IR and H$\alpha$, one can also see similar trends. Nevertheless, it is important to note that, at this time, lack of data causes strong limitations for a more detailed comparative analysis, especially in the cases of non-SN~IIn interacting SNe.

A further purpose of our current study has been to find any direct correlation between long-term X-ray/H$\alpha$ and mid-IR LC evolution, which has been poorly studied to date in the literature.
While comparative analyses of long-term X-ray, radio, H$\alpha$, optical, and near-IR SN LCs were published in several cases \citep[see, e.g.,][]{stritzinger12,fransson14,chandra15}, mid-IR LCs have been usually not (directly) involved in these comparisons.

A famous exception is the very nearby SN~1987A, for which the most complete long-term multiwavelength mid-IR/X-ray datasets were published \citep{bouchet06,dwek10,arendt16}. Only in this case, both very well-sampled mid-IR (from 3.6 to 24\,$\mu$m) and X-ray evolution of an SN can be followed during several thousand days, unfolding the ongoing interaction of the SN blast wave with a pre-existing dusty equatorial ring. In the listed papers, the IR-to-X-ray flux ratio (IRX) has been applied for investigating  the process of gas-grain collisions and cooling of the shocked gas.
As a shock sweeps though a medium with pre-existing dust, the IRX (at these mid-IR wavelengths) would be expected to decline as collisionally heated dust cools or is destroyed by sputtering. An epoch of dust formation in a CDS could act to increase the IRX, as could radiative processes (heating of the dust) in which case there is no direct physical interpretation of the IRX.
However, we note that the level of mid-IR activity measured in SN 1987A is far below the detection limit of {\it Spitzer} for the more distant extragalactic SNe, maybe because the CSM forms a ring rather than a shell in SN~1987A (however, even these low fluxes could be available for {\it JWST}).

Another example is the Type IIn SN~2005ip which also has been monitored for quite a long time \citep[$\sim 1000$--5000\,d;][]{fox20}. The mid-IR and (soft) X-ray \citep{katsuda14,smith17} LCs seem to follow similarly decreasing trends, in agreement with the scenario that the primary source of the observed mid-IR flux is warm dust radiatively heated by energetic photons emerging from CSM interaction, and that the shock may be finally reaching the outer extent of the dense CSM shell. The multiwavelength evolution,  including H$\alpha$, shows a shape that resembles that of an unabsorbed X-ray LC \citep[see][and also Fig.~\ref{fig:lc_IIn}]{smith17}. SN~2006jd, another SN~IIn, shows a quite similar behavior to that of SN~2005ip at various wavelengths; however, there is a lack of data after $\sim 2000$\,d (except in the mid-IR).

Based on the results of the first extended {\it Spitzer} survey of interacting SNe, \citet{fox13} also presented summarizing figures showing parallel mid-IR, X-ray, and optical ($R$-band) evolution of several SNe~IIn; however, most of those LCs consist of only very few points.


\begin{figure*}
\begin{center}
\includegraphics[width=.8\textwidth]{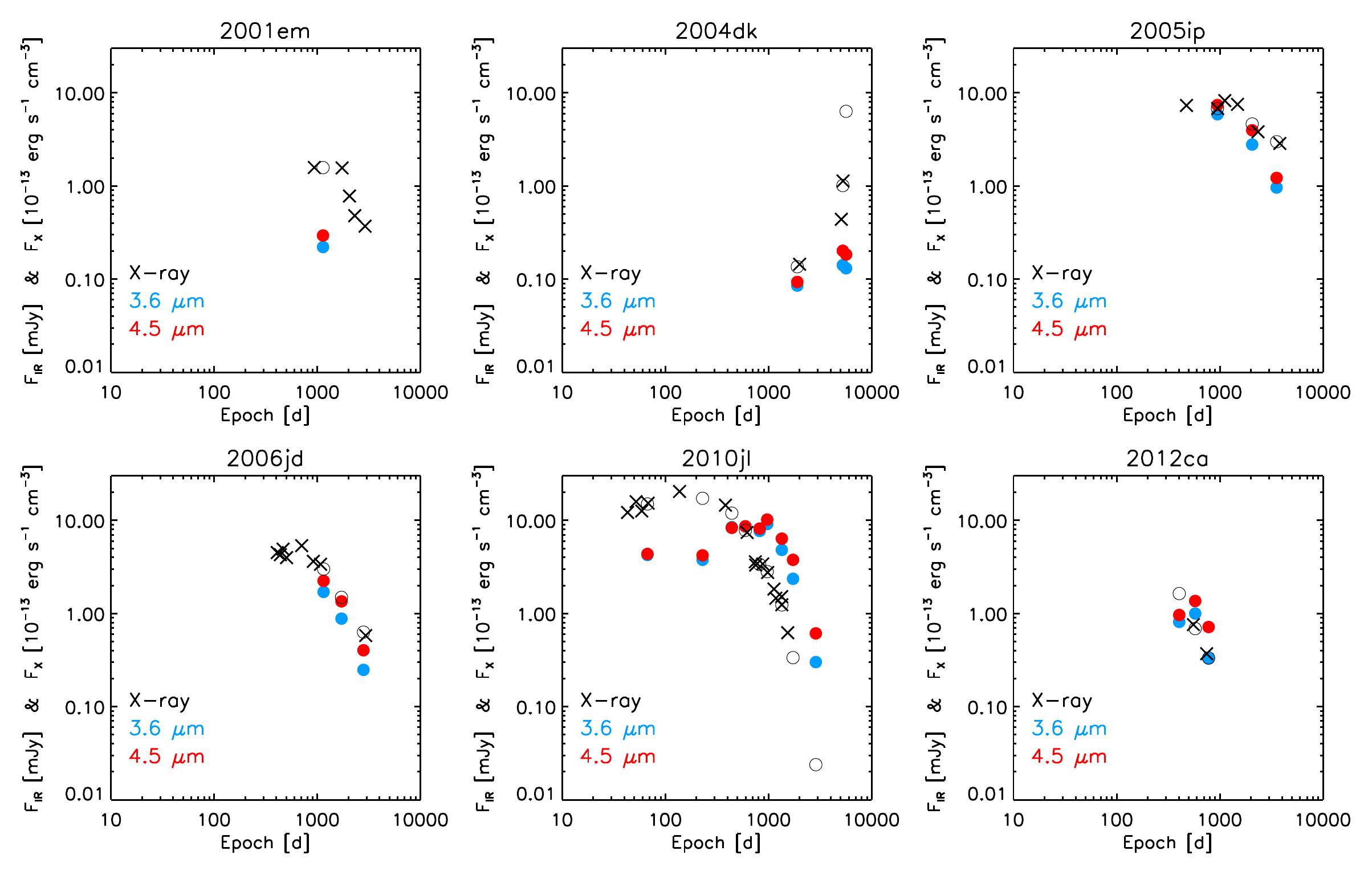}
\caption{Comparative evolution of X-ray and mid-IR fluxes of interacting SNe from our sample. Black circles indicate the X-ray fluxes interpolated to the epochs of the mid-IR observations.}
\label{fig:irx1}
\end{center}
\end{figure*}

\begin{figure*}
\begin{center}
\includegraphics[width=.8\textwidth]{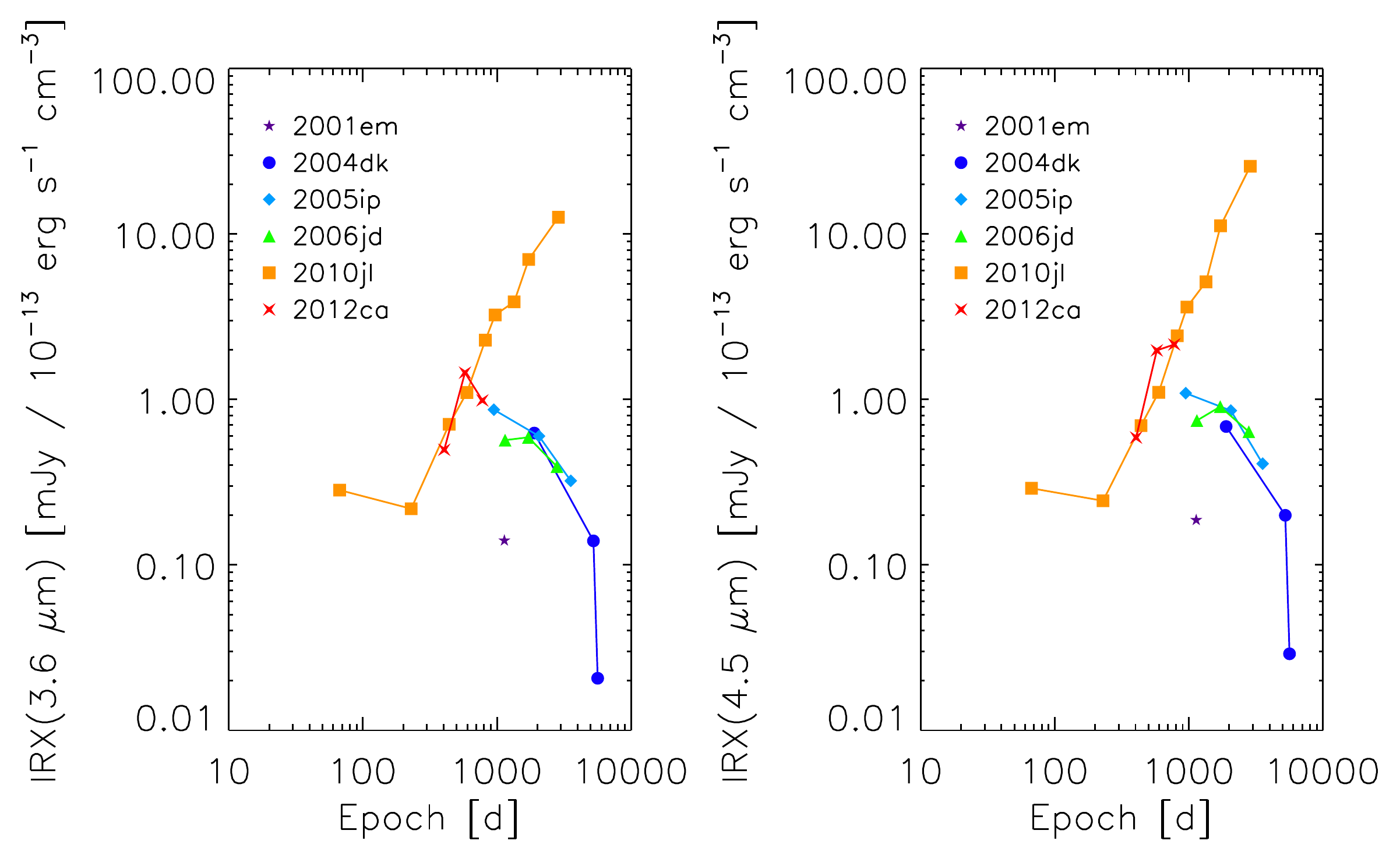}
\caption{IRX ratios for SNe presented in Figure~\ref{fig:irx1}.}
\label{fig:irx2}
\end{center}
\end{figure*}

In order to map similar correlations in our sample, we plotted the parallel evolution of all six SNe where both mid-IR and X-ray fluxes are available, and also calculated the IRX ratios for these objects (see Figures~\ref{fig:irx1} and \ref{fig:irx2}). 
Unfortunately, the amount and sampling of these data allow us to make only a preliminary analysis. Based on that, it seems that in SNe 2004dk, 2005ip, and 2006jd (expected to have radiatively heated pre-existing dust in their environments), all wavelengths trace each other \citep[as proposed by, e.g.,][]{fox11}.

While in most cases, comparative analysis of mid-IR and X-ray data was possible at epochs after $\sim 1000$\,d, it would be very interesting to see similar analyses at earlier phases. The only object with well-sampled mid-IR and X-ray data before $\sim 1000$~d is the Type IIn SN~2010jl (see Figs. \ref{fig:lc_IIn} and \ref{fig:lc_10jl}--\ref{fig:irx2}).
%
%
While SN 2010jl exhibits near-IR and mid-IR rebrightening and a quite long plateau after $\sim400$\,d, it is preceded by an X-ray/H$\alpha$ peak somewhere between 200 and 300\,d. As previously noted \citep[e.g.,][]{fransson14,gall14,sarangi18,bevan20}, an IR excess before $\sim 400$\,d may be the consequence of heating of pre-existing dust, while later-time mid-IR flux may originate either from new dust in the CDS and/or in the ejecta, or from CSM dust. \citet{sarangi18} presented a comparative analysis of mid-IR and X-ray LCs, and we complete that comparison here with the addition of an H$\alpha$ LC \citep[adopted from][]{fransson14}.

If we take into account general similarities between mid-IR/near-IR and between X-ray/H$\alpha$
(see above), we can also find some further cases where early-time multiwavelength evolution can be well traced.
Two other SNe~IIn, SN 2015da \citep{tartaglia20} and KISS15s \citep{kokubo19}, also show mid-IR LC evolution similar to that of SN~2010jl in the first $\sim 1000$\,d (up to the end of their datasets), probably going through similar circumstellar and dust-formation processes. In the case of SN~2015da, this is also strengthened by the parallel evolution of H$\alpha$ and mid-IR luminosities \citep[Fig.~\ref{fig:lc_10jl} ({\it right panel}), adapted from][]{tartaglia20}.

In SNe IIn 2005ip and 2006jd, one can see long plateaus in both the near-IR and X-ray/H$alpha$ ranges during the first several hundred days, in accordance with their later-time LCs also tracing each other (as described above).

There are also well-sampled near-IR/mid-IR and H$\alpha$ LCs of the Type II-P SNe 2004et and 2017eaw and of the Type II-P/L SN 2013ej (see Fig. \ref{fig:lc_I}). These LCs trace each other quite well, during both their declining and their small rebrightening phases (the latter ones can be seen at $\sim 800$--1000\,d). This can be explained well by CDS dust formation described in Section \ref{sec:res_mir_II-P}.

As a conclusion of this part, our preliminary studies show that comparative analysis of long-term multiwavelength (especially X-ray/H$\alpha$ vs. mid-IR) datasets is an essential (and maybe the only really useful) tool to differentiate between the existing dust forming/heating scenarios and, thus, to get a complete picture of the physical processes going on in the circumstellar environments of interacting SNe.
We plan to carry out a more detailed investigation of this topic, expecting near-future high-quality data (e.g., from approved {\it JWST} and associated programs).

%
%
%
%
%
%
%
%
%
%

\section{Conclusion} \label{sec:concl}

Here we have presented new {\it Spitzer} (3.6 and 4.5\,$\mu$m) photometry of 19 interacting SNe from various classes (18 of them observed during our LASTCHANCE survey), together with some nondetections. We also collected all previously published data for studying the long-term mid-IR evolution of these objects and for revealing the origin of this IR excess (which can be observed during thousands of days in certain cases).

We can draw some conclusions regarding every studied type of interacting SN. Assuming that the main source of mid-IR luminosity of SNe~IIn is pre-existing dust, the mid-IR brightness evolution of these objects further confirms the presence of extended dense CSM in their close environments. At the same time, in most cases, very late-time ($\gtrsim 2000$\,d) mid-IR fluxes continuously decrease, indicating that CSM interaction ultimately weakens, possibly at a predictable rate.

The detected, homogeneous mid-IR evolution may hint that SNe~Ia-CSM explode in very similar environments surrounded by dense but less-extended CSM than SNe~IIn do. 
While further data are necessary to reveal the true nature of SN~Ia-CSM explosions, the IR presented mid-IR LCs may hint that their progenitors form a homogeneous class (contrary to SNe~IIn), which also can support their thermonuclear origin.
Mid-IR data of the stripped-envelope SNe in our sample support the idea that there can be detached CSM shells around some of these objects.  Delayed CSM interaction may not begin for years, or even more than a decade, after explosion.
Mid-IR properties of intermediate-luminosity interacting transients, just as their optical ones, are quite heterogeneous. These results support the assumptions that either the progenitors, explosion mechanisms, or environmental circumstances of these objects may be quite different.

Comparison of long-term mid-IR LCs with those obtained in other wavelength ranges (especially X-ray, H$\alpha$, and near-IR) can be an efficient method to reveal otherwise hidden details of the connection of CSM interaction and formation/heating of ambient dust. Our analysis of multiwavelength datasets of SNe, especially regarding mid-IR vs. X-ray luminosity evolution, supports previous results: some interacting SNe (such as SNe~2004dk, 2005ip, and 2006jd) seem to have large amounts of radiatively heated, slowly cooling pre-existing dust in their environments, while others (such as SN 2010jl, but probably also SN 2015da or KISS 15s) exhibit signs of dust formation after explosion.

Our current study supports the idea that long-term mid-IR follow-up observations can play a key role for a better understanding of both pre- and post-explosion processes in exploding stars and in their environments. While {\it Spitzer}, the most effective tool that has been used for this purpose, is not available anymore, we have good prospects for the near future. Expected observations with {\it JWST} will have unique sensitivity to cooler dust grains at wavelengths $>4.5$\,$\mu$m and faint emission from SNe even years after explosion that would have gone undetected by {\it Spitzer} or any other mid-IR spacecraft. We also showed that, in dusty SNe, near-IR ($K$-band) LC evolution is quite similar to what we can see in the mid-IR. Thus, either ground-based or (upcoming) near-IR space telescopes (e.g., {\it NGRST}) can be effectively used for monitoring warm SN dust evolution, as well as in selecting appropriate targets for {\it JWST} and for other top-class IR telescopes in the future.\\

\vspace{3mm}
\acknowledgements
This work is based on observations made with the {\it Spitzer Space Telescope}, which was operated by the Jet Propulsion Laboratory, California Institute of Technology, under a contract with the National Aeronautics and Space Administration (NASA). Support for this work was provided by NASA through an award issued by JPL/Caltech. This research has made use of the NASA/IPAC Infrared Science Archive and NASA/IPAC Extragalactic Database (NED), which are both operated by JPL/Caltech, under contract with NASA; the SIMBAD database, operated at CDS, Strasbourg, France; the Supernova X-Ray Database (SNaX) funded by NASA Astrophysics Data Analysis program grant NNX14AR63G awarded to the University of Chicago; and the Open Supernova Catalog. We acknowledge the availability of the SAO/NASA Astrophysical Data System (ADS) services.
This project has been supported by the GINOP-2-3-2-15-2016-00033 project of the National Research, Development and Innovation Office of Hungary (NKFIH) funded by the European Union, and by NKFIH/OTKA FK-134432 grant.
T.S. is supported by the J\'anos Bolyai Research Scholarship of the Hungarian Academy of Sciences and by the New National Excellence Program (UNKP-20-5) of the Ministry for Innovation and Technology of Hungary from the source of the National Research, Development and Innovation Fund.
A.V.F. is grateful for support from the U.C. Berkeley Miller Institute for Basic Research in Science (in which he is a Senior Miller Fellow), the TABASGO Foundation, the Christopher R. Redlich Fund, and many other individual donors.

\vspace{2mm}
\facilities{Spitzer(IRAC)}

\software{IRAF}

\vspace{3mm}
\bibliography{main}{}
\bibliographystyle{aasjournal}




\end{document}